# Depth-resolving the redox compensation mechanism in $Li_xNiO_2$

*Roberto Fantin [a], Thibaut Jousseaume [b], Raphael Ramos [a], Gauthier Lefevre [a], Ambroise Van Roekeghem [a], Jean-Pascal Rueff [c], Anass Benayad [a*]*

[a] Univ. Grenoble Alpes, CEA-LITEN, Grenoble, 38054, France

[b] Univ. Grenoble Alpes, CEA-IRIG, Grenoble, 38054, Cedex 9, France

[c] Synchrotron SOLEIL, L'Orme des Merisiers, Saint-Aubin, BP 48, 91192 Gif-sur-Yvette Cedex, France

AUTHOR INFORMATION

**Corresponding Author**

* anass.benayad@cea.fr

# ABSTRACT

The performances of lithium-ion batteries are set by the electrodes materials capacity to exchange lithium ions and electrons faster and reversibly. To this goal Ni-rich layered metal oxides, especially $LiNiO_2$, are attractive electrode candidate to achieve both high voltage and capacities. Despite its attractiveness, several drawbacks for its industrialization are related to different form of surface and bulk instabilities. These instabilities are due to redox process involving the charge transfer between cations and anions. Therefore, a fundamental understanding based on further experimental evidence is required to resolve of charge transfer between the cation and anion from the surface to the bulk in $LiNiO_2$. Herein, we resolve the role of nickel and oxygen in the charge compensation process in $Li_xNiO_2$ electrodes from the extreme surface down to 30 nm by energy-dependent core-level HAXPES supported by *ab initio* simulation. We emphasize the central role of oxygen in the bulk charge compensation mechanism from $LiNiO_2$ to $NiO_2$ due to the negative charge transfer and bond/charge-disproportionation characters of $LiNiO_2$. This bulk behavior is in turn responsible for surface deoxygenation and nickel reduction upon delithiation.

# INTRODUCTION

Lithium-ion batteries energy density and specific capacity performances are set by the electrodes materials capacity to exchange lithium ions and electrons faster and reversibly. Understanding the mechanism behind these processes sets the direction for designing high density positive electrode material with good structural and chemical stability upon different cycling regimes. Toward this goal, the interest in $LiNiO_2$ has been renewed due to the progressive increase of the nickel content in Li-ion battery positive electrode materials $LiNi_xMn_yCo_zO_2$ (x+y+z=1, NMC)[1]. This strategy, coupled with the raise of the upper cut-off voltage towards the maximum theoretical capacity, aims to increase batteries energy density[2,3]. However, at high $Li^+$ de-intercalation levels, Ni-rich NMCs suffer from bulk and surface degradation[4–9], gas release[10–12], and particle cracking[13–15]. Reaching deeper states of charge requires a precise understanding of the charge compensation mechanism sustaining $Li^+$ de-/intercalation[16].

$LiNiO_2$ is nominally a Jahn-Teller-active $3d^7$ material although no cooperative but only local distortions of the $NiO_6$ octahedra were observed[17–20]. Following previous studies for Li-doped NiO[21–23] and rare earths $d^7$ nickelates[24–26], $LiNiO_2$ is nowadays proposed as a negative charge transfer material with an average $|d^8\mathbf{L}\rangle$ local electronic structure ($\mathbf{L}$ being a hole delocalized in the O 2p ligand orbitals), resulted by bond and charge disproportionation[27,28]. First-principles simulations suggested a negative charge transfer also for $NiO_2$ [29], in agreement with recent evidence of double ligand holes ($\left|d^8\boldsymbol{L}^2\right\rangle$) in delithiated $Li_xNiO_2$[30].

The presence of oxygen holes in both stoichiometric and Li-rich layered oxides is assumed by O K-edge Resonant Inelastic X-ray Spectroscopy (RIXS) by the appearance of a peak at emission and excitation energies of ~523.5 and ~531 eV, respectively [31–37]. For $Li_xNiO_2$, the appearance of such feature at V > 4.3 V vs $Li^+/Li$ was referred to a lattice oxygen oxidation from $O^{2-}$ to $O^{n-}$ ($1<n<2$) which is triggered after near-full oxidation of $Ni^{3+}$ to $Ni^{4+}$ in their ionic low-spin configurations as $t_{2g}^6e_g^1$ and $t_{2g}^6e_g^0$, respectively [33]. However, bulk oxygen redox is typically related to O 2p lone pairs that are present in Li-rich compounds but not in $LiNiO_2$ [31,32,38]. In fact, a recent O K-edge RIXS study disputed this picture and related the observed features to the strong Ni 3d - O 2p hybridization instead, as already highlighted by X-ray Absorption spectroscopy (XAS) studies [21,22,32,39–42], supporting the negative charge transfer process for $LiNiO_2$ [32]. Therefore, XAS and RIXS experimental insights on oxygen redox call for parallel investigations by other techniques and highlight the need of coupling with atomistic simulations, towards unification of the theories developed for stoichiometric and Li-rich oxides [31,32]. Moreover, a nanometer-scale depth resolution is also necessary to resolve the charge compensation mechanism from the surface towards the bulk.

High surface resolution and sensitivity to local electronic structures are characteristics of X-ray photoelectron spectroscopy (XPS). Most XPS studies of positive electrode materials focused on the characterization of coatings, surface stability, and the solid electrolyte interphases at the positive electrode (pSEI)[7,43–47]. Only few investigations on the electronic structure, limited to $LiCoO_2$ and NMC compositions and no study for $LiNiO_2$, are present in the literature[48–52], mainly employing soft X-rays with a typical depth sensitivity of less than 5 nm. To overcome this limit, synchrotron-based hard X-ray photoelectron spectroscopy (HAXPES) received large attention[53–55]. However, unravelling bulk and surface contributions in HAXPES spectra is nontrivial as highlighted for the O 1s spectra of Li-rich oxides, where the same peak at binding energy of ~530 eV was assigned to either bulk oxidation[53] or surface degradation[56]. Such issues can be addressed by depth-resolved quantitative analysis, nowadays available with lab-based HAXPES equipment [57–59].

Numerous efforts have been dedicated to determining Ni oxidation states through Ni 2p XPS analysis, through the satellite structures related to Ni 2p - Ni 3d multiplets and Ni 3d - O 2p charge transfer screening processes[60]. This task presents several challenges and no standardized method is present in the literature. Indeed, while in the most simple cases the oxidation can be determined by chemical shifts of single peaks eventually quantified by peak fitting, this method is not valid for late 3d transition metals[60]. Yet, single peak areas of the main line are still incorrectly used to quantify the $Ni^{II}/Ni^{III}$ to reveal the redox process in battery-related studies[61–68]. Other methods were based on peak templates based on Ni ions multiplets[69,70], satellite analysis by peak fitting[71], and reference-based peak fitting[72]. A significant step forward would consist of a direct coupling with atomistic simulations as proposed in this work.

This study has therefore a dual objective: (1) following the evolution of Li, Ni and O local electronic structures upon delithiation at increasing depths in the nanometer scale and (2) understanding the redox compensation mechanism in $Li_xNiO_2$. The surface and bulk contributions to the O 1s and Ni 2p core-level spectra were discriminated by performing nondestructive XPS and HAXPES depth profiles of $Li_xNiO_2$ electrodes at different states of charge. The qualitative and quantitative analysis revealed the presence of a ~10 nm thick surface layer in which Ni and O are in a reduced state, while the bulk is progressively oxidized, consistently with X-ray diffraction (XRD), Raman spectroscopy, and energy electron loss spectroscopy (EELS) characterization. The participation of O and Ni to the redox reaction was evaluated by analyzing the evolution of O 1s and Ni $2p_{3/2}$ HAXPES satellite peaks using *ab initio* calculations based on density functional theory (DFT), constrained random phase approximation (cRPA), and cluster model calculations. The results indicate that the charge compensation from $LiNiO_2$ to $NiO_2$ is mainly carried out by the O 2p states, highlighting the negative charge transfer and bond/charge-disproportionation characters of $LiNiO_2$.

# RESULTS

## Structural characterization

Positive electrodes consisting of $LiNiO_2$ active material, conductive carbon additive and binder were provided by BASF. $Li_xNiO_2$ *ex-situ* samples were prepared by electrochemical $Li^+$ deintercalation in Li-ion coin-cells with graphite/LP57/$LiNiO_2$ configuration. The electrochemical protocol consisted of three formation cycles between 2.5 and 4.2 V followed by slow constant current constant voltage (CCCV) charging up to selected upper cutoff voltages of 3.8, 4.2, and 4.8 V and a final relaxation step. At the end of the charge, $Li_xNiO_2$ samples with $x \approx 0.5$, 0.1, and 0.01 were obtained (**Supplementary Fig. 1**). A sample without the last charge step ($x \approx 0.8$) was also included in the analysis. The crystal structure of all samples was characterized by synchrotron X-ray diffraction (**Supplementary Fig. 2** and **Supplementary Tab. 1**), confirming bulk delithiation in agreement with the typical phase diagram of $Li_xNiO_2$ [73].

## Non-destructive depth profile approach for *ex-situ* $Li_xNiO_2$ samples

The *ex-situ* $Li_xNiO_2$ samples were investigated by lab-based and synchrotron XPS/HAXPES. Core level spectra were measured with five increasing sampling depths according to the increasing photon energies, namely 1.5*, 2.3, 5.4*, 5.4 and 9.5 keV (star symbol for laboratory X-ray sources). Laboratory-based and synchrotron HAXPES measurements at 5.4 keV gave different sampling depths because of the take-off angles of 45° and 80°, respectively. According to the inelastic mean free path (IMFP) calculations (**Supplementary Tab. 2**), the combination of all experiments uniformly allowed scanning the first 30 nm (**Fig. 1**). A preliminary study indicated that synchrotron beam damage mostly affected the chemical structure of the pSEI, leaving the buried layered oxide, i.e. the target of our analyses, nearly unchanged [74].

## Tracking the surface-to-bulk distribution of $Li^+$ ions

**Fig. 2a-e** show the Li 1s spectra, normalized to their background intensity for direct comparison. The Li 1s core level could not be measured by lab-based HAXPES due to low cross-section and X-ray flux. Regarding the pSEI part, $Li^+$ is expected in different chemical environments that can be distinguished by their binding energy[45,75]. In our case, fitting with two pseudo-Voigt peaks gave more consistent results.

The two peaks were assigned to $Li^+$ in the oxide lattice structure ($Li_{latt}$, ~54 eV) and in the surface species ($Li_{surf}$, ~56 eV). The latter contribution, due to a native surface film for the pristine material or to the pSEI for the cycled electrodes, includes LiF, $Li_2CO_3$ and LiOH as confirmed by C1s and F1s analysis (**Supplementary Note 1**). At high states of charges (**Fig. 2d,e**), for which the Li concentration in the $Li_xNiO_2$ is expected to be below the detection limit of HAXPES, the $Li_{latt}$ peak was still measured closer to the surface. This suggests the presence of a few nm surface layer with a $Li_{latt}$ concentration higher than in the bulk, interpreted as a $Li^+$ ion trapping in the NiO-like rocksalt layer due to surface reconstruction of layered oxides [5,8,9,14].

## Decoupling surface and bulk oxygen states

The O 1s spectra are shown in **Fig. 3a-e**. Four pseudo-voigt peaks were used to fit the data in the 528-535 eV region. These peaks were assigned to oxygen in layered lattice ($O_{latt}$, 528 eV), in the surface layer ($O_{surf}$, 530 eV), and in the surface species ($O_{carb/hydr}$, 531.5 and $O_{org}$ 533 eV), in line with previous reports for layered oxides[4,45,55,56]. The latter contributions starkly decrease with increasing probing depth, indicating a very thin pSEI layer, in agreement with the 1-2 nm calculated thickness (**Supplementary Note 2**).

As evidenced by the bulk-sensitive HAXPES spectra, the peak at 530 eV, often related to oxygen oxidation, is present even at the pristine state. Its area percentage with respect to $O_{latt}$ is shown in **Fig. 3f** for each sample as a function of the probing depth, showing slight voltage dependence but a tendency to decrease towards the bulk. Therefore, we assign this peak to surface and bulk defects and degradation rather than oxygen oxidation upon charging.

This is supported by the O/Ni at% ratio estimated by lab-based XPS and HAXPES using measured relative sensitivity factors for 1.5 and 5.4 KeV X-ray energy (**Fig. 3g,h**). Based on O 1s and Ni 2p peak fitting (**Supplementary Note 3**), taking the whole Ni 2p area with respect to the $O_{latt}$ peak gives good agreement with attended composition in the case of reference samples NiO and $Ni(OH)_2$ but not in the case of real electrodes ("as ref" values). Even adding the higher binding energy O 1s peak in the calculation ("all-in"), e.g. assuming this contribution is related to bulk oxidation, the O/Ni ratio is <2, indicating surface sub-stoichiometry. When Ni 2p and O 1s surface and bulk contributions ("surf" and "bulk") are separated, a better agreement with the expected ratio for the bulk layered structure ($Li_xNiO_2$, O/Ni~2) and surface layer (NiO-like, O/Ni ~ 1) is obtained. This result highlights the strength of XPS and HAXPES quantitative analysis to support peak fitting models.

### Delithiation induces an increase in O 1s plasmonic excitations

Although we excluded that the 530 eV peak is related to oxygen oxidation, the high-energy satellites observed in O 1s spectra could be a signature of its redox activity. As highlighted in the zoomed spectra measured at 9.5 keV, in the pristine material only a weak satellite at ~541 eV can be distinguished, whereas upon charging two peaks at 537 and 540 arise. Similar satellites were observed in Li-rich NMCs and other materials suspected of so-called anionic redox, although their origin is not yet understood[53,54]. Therefore, clarifying this would help understanding the role of O 2p states in the charge compensation mechanism in these materials.

We evidence here that the O 1s satellites are related to plasmonic excitations process, as resulted by constrained random phase approximation (cRPA) calculations of the dynamically screened Coulomb interaction $U(\omega)$[76]. Indeed, plasmonic excitations can be recognized by a divergence of the real part of the $U(\omega)$ curve at the plasmon frequency $\omega_p$ [77]. As shown in **Fig. 3i-k**, plasmons were observed at $\omega_p$ = 8.5 eV for $LiNiO_2$ and $\omega_p$ = 8.5 and 11 eV for $NiO_2$. While a good quantitative agreement is obtained for $NiO_2$ even at DFT level, the same cannot be said for $LiNiO_2$. An explanation for this mismatch could be the limitations of DFT for describing the electronic structure of $LiNiO_2$ -- wrong prediction of a metallic state without local disproportionation -- (**Supplementary Note 4**); otherwise, the satellite peak measured for $LiNiO_2$ could be due to other physical processes e.g. local charge transfer.

To understand the origin of the strong plasmonic excitations in the O 1s HAXPES spectrum for $NiO_2$, we compared the results obtained from model calculations including or not the O 2p to Ni 3d screening channels, shown in **Fig. 3j,k** as $U_{d\text{-}dp}$ and $U_{dp\text{-}dp}$, respectively. A schematic representation highlighting the difference between the two models is given in the **Supplementary Fig. 10**. Since no divergence was found in the latter case, it becomes clear that these plasmons are related to p-d charge transfer-like transitions, indicating an increase of such screening channels going from $LiNiO_2$ to $NiO_2$.

## Decomposition of Ni $2p_{3/2}$ into surface and bulk contributions

The normalized Ni $2p_{3/2}$ spectra are shown in **Fig. 4a-e**. Looking to the surface sensitive spectra (top row), only a lowering of the low-binding energy shoulder at 853 eV can be detected along the delithiated series of materials, while the satellite structure at 860-865 eV is nearly unchanged. On the contrary, the bulk-sensitive HAXPES spectra of delithiated $Li_xNiO_2$ show clear differences upon delithiation, namely a sharper main line at 855 eV and a satellite structure shifting towards higher binding energies, which is typically associated to higher oxidation state [78]. Therefore, it is clear that a surface layer with Ni in reduced state formed upon cycling while bulk Ni oxidation could be observed within the first 30 nm probed by multiple HAXPES solicitation. The surface reduction of Ni matches with the increase of $O_{surf}$ peak intensity in O 1s spectra and the presence of surface $Li_{latt}$, leading to a description of this surface layer as $Li_xNi_{1-x}O$.

Due to the above-mentioned complex structure of Ni $2p_{3/2}$ core levels spectra, a reference-based peak fitting was performed to distinguish surface and bulk Ni local electronic states. The approach was based on three internal references: bulk sensitive spectra of pristine and deeply delithiated samples and surface sensitive XPS spectra for the discharged state samples. We refer to them as $Ni^{III}_{bulk}$, $Ni^{IV}_{bulk}$, and $Ni^{II}_{surf}$ contributions, respectively, indicated by star symbols in **Fig. 4a,b,e**. Such approximations result from the difficulties to identify these contributions due to the integrated nature of the depth sensitivity in photoemission experiments. To partially overcome this problem, the contribution of $Ni^{II}_{surf}$ was included to fit the experimental HAXPES spectra of $Li_{0.01}NiO_2$ to define the $Ni^{IV}_{bulk}$ line shape.

The decomposition of Ni $2p_{3/2}$ into surface and bulk contributions allowed the O/Ni quantification discussed above (**Fig. 3f,g**) and supported the 10 nm thickness estimated for surface $Li_xNi_{1-x}O$ layer, which agrees well with scanning transmission electron microscopy thickness estimation (**Supplementary Fig. 7**). However, the use of reference spectra for peak fitting does not to give insights about the actual electronic structures beneath the core level spectra, leading us to further analysis based on *ab initio* simulations.

## Unraveling the Ni core-level final states

First, the nature of Ni core level satellites was determined by looking at the Ni 1s spectra. In fact, while both (non-)local charge transfer and intra-atomic core-valence screening channels can contribute to the different final states upon photoemission in a transition metal oxide, their relative weight is core-level dependent [60,79,80]. In case of deep core levels such as Ni 1s, core-valence multiplet effects are negligible, allowing to isolate the charge-transfer processes [80]. The Ni $2p_{3/2}$ and Ni 1s core level spectra for the pristine (top) and deeply delithiated (bottom) samples are compared in **Fig. 4f**. For the former, the strong similarities indicate that the Ni $2p_{3/2}$ satellites have a dominant charge-transfer character. For the cycled sample, the disagreement between Ni 1s and Ni 2p measured at 9.5 keV is explained by the different depth probed by the two measures due to the large difference in binding energy, as proved by comparing the Ni 1s spectra with the Ni 2p spectra acquired at 2.3 keV, with a similar IMFP of ~2-3 nm.

## Understanding Ni local electronic structure

Once demonstrated the need of using charge-transfer models to interpret the Ni 2p XPS spectra of nickelates, we performed *ab initio* based cluster model calculations to simulate the Ni 2p spectra and study the charge compensation mechanism in $Li_xNiO_2$[81,82]. This method has been successful for studying several transition metal oxides [83–85] and was recently employed for determining the charge transfer in the $LiCoO_2$-$CoO_2$ system [52]. Due to the difficulty in fitting the experimental spectra, we relied on *ab initio* DFT and cRPA calculations for hopping and screened Coulomb interactions, respectively. **Fig. 4g** shows the ground state occupation probabilities for the Ni 3d states in NiO, $LiNiO_2$, and $NiO_2$, showing that they all have a dominant $d^8$ character. This supports the idea that the charge compensation mechanism is based on O 2p to Ni 3d charge-transfer with a major role played by O 2p states, in agreement with previous studies [21,27,29,30]. The corresponding simulated and experimental HAXPES spectra of NiO, $LiNiO_2$, and $NiO_2$ are shown in **Fig. 4h**, where the experimental Ni $2p_{3/2}$ spectra peak fitting and the partial simulations were included to show the character of the satellite peaks.

The simulated spectra for NiO agrees overall well with the experiment except for the lowest energy peak. This peak was assigned to non-local screening (NLS) processes, which are missed in single cluster simulations[23,80,86–88]. The high-energy satellite structure is due to unscreened multiplet $|d^8\rangle$ states, as shown by the partial simulated spectra.

The simulation for $LiNiO_2$ led to a single satellite peak, in contrast to the broad structure of the experimental spectra. Since the whole satellite structure was related to charge-transfer screening (**Fig. 4f**), typically well treated by cluster model calculations, we propose that this mismatch can be explained by considering the charge and bond disproportionation model of Foyevtsova et al. [27]. Indeed, the three cluster simulations for $NiO_2$, $LiNiO_2$ and NiO can be related to the small, medium, and large clusters with $|d^8\boldsymbol{L}^2\rangle$ , $|d^8\mathbf{L}\rangle$, and $|d^8\rangle$ electronic structures presented in ref [27]. By superposing the Ni 2p cluster simulations for NiO, $LiNiO_2$, and $NiO_2$, the satellite structure of $LiNiO_2$ is overall well described.

Delithiation from $LiNiO_2$ to $NiO_2$ led to a Ni $2p_{3/2}$ spectra with a spectral weight transfer to the satellite peak at high energies identified as $|d^8\boldsymbol{L}^2\rangle$ by the partial spectra, and less disproportionation character, considering the overlapping surface contributions ($Ni_{surf}$). The lifting of charge disproportionation upon delithiation is correlated to the increase of O 1s plasmon satellites (**Fig. 3m**), which have a strongly delocalized character and therefore might be suppressed in the case of $LiNiO_2$ disordered electronic structure.

Such ordering of the electronic structure upon delithiation fits well with the evolution of the Ni-O vibrational modes perpendicular ($E_g$) and parallel ($A_{1g}$) to the c-axis as measured by Raman spectroscopy at the single particle scale level (**Fig. 3i**)[89,90]. In fact, while the $A_{1g}$ mode relates to overall $NiO_6$ cluster expansion/contraction, only the $E_g$ mode induces JT-like octahedra distortion[91]. Bond disproportionation in pristine $LiNiO_2$ would therefore perturb the latter mode while enhancing the former, explaining both the broadness and low intensity of the $E_g$ peak and the opposite for $A_{1g}$ one. Upon delithiation, the lifting of the bond-disproportionation in favor of a more homogeneous and strongly covalent electronic structure is reflected by the narrowing and increase of the $E_g$ mode.

# DISCUSSION

The energy-dependent HAXPES approach allowed to relate electronic structure changes with high sensitivity to local and collective charge transfer processes, resolved across the first 30 nm from the uppermost surface. The crucial role played by oxygen, strongly related to O 2p - Ni 3d charge transfer, was highlighted by Ni 2p and O 1s satellites structure evolution. While the origin of O 1s satellites as plasmonic excitations was revealed for the first time, further investigation on the nature of these oscillations is expected to shade more light on $Li^+$ diffusion mechanism and structural evolution of the layered host.

However, we proved that the O 1s peak at 530 eV is not related to oxygen redox but to intrinsic surface defects in the material and surface degradation upon cycling, which consists of oxygen loss and nickel reduction. Therefore, the development of better positive electrode materials must account for the two-faced role of oxygen: on the one hand, it is essential in compensating the (de-)lithiation process; on the other, it leads to surface instabilities and degradation. Suppressing the latter by surface engineering while enhancing the former via accurate bulk doping are therefore promising pathways to increment materials capacity and stability at high voltage.

# METHODS

## Materials

$LiNiO_2$ pristine electrodes with a loading of 3 $mAh/cm^2$ were provided by BASF and consist of 94 wt% $LiNiO_2$ (BASF), 3 wt% C65 carbon black (TIMCAL) and 3 wt% Solef5130 polyvinylidene fluoride (PVDF, Solvay). The three components were also analyzed separately as bare powders. Commercial-grade NiO and $Ni(OH)_2$ powders (Sigma Aldrich) were used with no further treatment.

## Electrochemical methods

CR2032 lithium-ion cells were assembled in an Ar-filled glovebox as follows, in order of assembling: 15 mm diameter graphite anode (CIDETEC), 16 mm Celgard separator, 50 μL of a 1 M $LiPF_6$ solution in ethylene carbonate (EC) and ethyl methyl carbonate (EMC) with weight ratio 3:7 (LP57 electrolyte, Sigma Aldrich), 14 mm diameter LNO electrode. All cells were tested with a VMP-300 (Biologic) potentiostat at room temperature with the following protocol: (1) 10 hours at open-circuit voltage (OCV), (2) three galvanostatic cycles between 3.0 and 4.2 V at C/10 (1C = 225 mA/g) including a 5h constant voltage step at 4.2 V, (3) galvanostatic charge to 3.8, 4.2, and 4.8 V upper cut off voltage at C/20, (4) constant voltage until the current was lower than C/100, (5) final relaxation step at OCV. The cycled electrodes were recovered from disassembled cells and washed with dimethyl carbonate (DMC) for about 3 min to remove soluble salts from the surface. Cell disassembling, electrodes washing and storage were all performed in an Ar-filled glovebox.

**X-ray Diffraction**

Synchrotron X-ray diffraction measurements were performed at beamline BM32, at the European Synchrotron Radiation Facility using a beam energy set to 27 keV. The beam had a size of about 150μm × 550μm and a flux of about $3 \cdot 10^{10}$ph.s$^{-1}$. The XRD patterns of the *ex situ* samples were recorded in transmission geometry (Debye-Scherrer) with a moveable 2D CdTe detector, calibrated and integrated into intensity *vs.* angle 1D-patterns using the pyFAI multigeometry module [92]. The setup and the data integration process provided an angular resolution of $7.8 \cdot 10^{-3}$° in the range [2.5°, 32.7°]. The cycled electrodes of LNO were covered by polyimide tape to limit air contamination before the measurement. The covered electrodes were brought to the beamline in pouch casings sealed in water free (<1ppm) argon atmosphere.

The patterns were analysed with the Fullprof software [93] by Rietveld refinement. Lattice parameters, oxygen position, Ni/Li anti-site default, thermal movement of the Ni in the layer and strain of the H3 phase were refined. The peaks were fitted using pseudo-Voigt functions.

## Scanning electron microscopy

Scanning electron microscopy (SEM) images were collected using a MEB-LEO microscopy using a primary electron beam of 5 KV acceleration voltage.

## Scanning transmission electron microscopy

STEM images and EELS spectra were acquired on a probe Cs-corrected TFS Titan Themis microscope operating at 200 kV. EELS spectra were collected in spectrum imaging mode with a Gatan GIF Quantum electron spectrometer using a dispersion of 0.25 eV per channel and a 2.5 mm aperture. Concerning the sample preparation, powders were ground and deposited on a lacey carbon coated grid. For the observations, grids were transferred from the glovebox to the microscope using a vacuum transfer holder in order to protect the sample from air exposure.

## Raman spectroscopy

Raman spectroscopy analyses were performed using a Renishaw inVia Raman microscope in backscattering configuration with 532 nm laser excitation and low power ($< 0.1$ mW/$\mu m^2$ on the sample) to prevent sample heating or degradation. The electrodes were directly transferred – without preparation - and sealed inside a custom dedicated airtight optical sample holder in an Ar-filled glovebox to avoid exposure to lab atmosphere.

The lateral spatial resolution of microscope (~1.5 µm) allowed laser focusing on individual $Li_xNiO_2$ spherical aggregates. Multiple aggregates were systematically analyzed to assess homogeneity of both pristine electrode materials and electrochemical charging process.

The depth probed by Raman analysis depends on the optical properties of the electrodes, hence on their state of charge. The most opaque electrodes are the pristine ones, for which a penetration depth in the 100 to 200 nm range can be estimated at 532 nm based on reported extinction coefficients [94,95] in agreement with previous estimates for Ni-rich layered oxides [90]. Laser penetration depth is thus comparable to the radius of primary $LiNiO_2$ particles, and it increases for cycled electrodes [95].

### X-ray photoelectron spectroscopy

Laboratory XPS and HAXPES measurements were performed with a QUANTES spectrometer (ULVAC-PHI) equipped with a co-localized Al kα (1486.6 eV, 25 W) and Cr kα (5414.9 eV, 50 W) dual X-ray source. The take-off angle for photoelectron detection was 45°. All measurements were performed under ultrahigh vacuum conditions. High-resolution spectra were acquired with a pass energy of 55 eV for both energy sources, corresponding to an energy resolution of about 0.71 and 0.93 eV for Al kα and Cr kα, as estimated from the FHWM of the Ag $3d_{5/2}$ of a reference Ag sample.

The electrode samples were prepared in Ar-filled glovebox. Cycled electrodes recovered from disassembled coin cells were washed in few milliliters of dimethyl carbonate (DMC) for 1 min. After mechanical scratching with ceramic knife, the electrode samples were mounted to the PHI sample holder using copper double tape. The transfer from the glovebox to the Quantes equipment was carried out using a dedicated transfer vessel.

To minimize the X-ray beam damage effect, all measurements were taken using an X-ray beam spot size of 100 μm diameter for both sources scanning an area of 500 x 500 $um^2$ defined by secondary X-ray imaging of the sample. Further comments on the beam damage effect for these samples can be found elsewhere [74]. Because of the low conductivity and absence of beam damage for the bare powders, dual electron and Ar-ion charge neutralization was enabled and the measurements were carried out using a single X-ray beam spot size of 100 µm diameter. The reproducibility was checked by comparing lab-based measurements for two electrodes prepared with the same electrochemical protocol.

Synchrotron HAXPES was performed in the Galaxies beamline of Soleil using monochromatic X-rays with three photon energies: 2300, 5400, and 9500 eV. The incident and take-off angles were 10° and 80°, respectively. The X-ray beam spot was 20x80 $um^2$ (oval shaped). A different spot of the sample was used for each X-ray energy. The pass energy was 200 eV corresponding to an energy resolution from 0.3 eV for the measures at 2300 and 9500 eV and 1.0 eV for the one at 5400 eV. No charge calibration was employed. A cut of the same electrodes measured by lab-based XPS/HAXPES were transferred using double thermally sealed pouches to prevent air contamination and also prepared in an Ar-filled glovebox using copper double tape.

To quantify the different sampling depths of XPS and HAXPES techniques, the inelastic mean free paths (IMFPs) were calculated with the Tanuma-Powell-Penn method by the Quases-IMFP-TPP2M software [96]. The values are shown in the **Supplementary Table 2** and were used to calculate the sampling depth as $3\lambda \sin\vartheta$, with $\vartheta$ the take-off angle of 45° and 80° for lab-based and synchrotron experiment, respectively.

All spectra were analyzed with the CasaXPS software. The relative sensitivity factor (RSF) for both X-ray sources were exported from the MultiPak library to take into account the transmission function of the spectrometer. For insulating reference samples ($Li_2CO_3$, NiO, $Ni(OH)_2$, PVDF), binding energy charge correction was performed to the C 1s peak at 285 eV. No charge correction as found to be necessary for the electrodes. For the Li 1s and O 1s spectra, peak decomposition was performed with Pseudo-Voigt functions having 30% and 70% of Lorentzian and Gaussian weight for each component and iterated Shirley backgrounds.

The reference-based peak fitting for the Ni $2p_{3/2}$ core level spectra was carried out by defining experimental-based lineshapes within CasaXPS software. Three components were defined by peak decomposition of the spectra with a model templates derived from NiO. The signature of bulk discharged $LiNiO_2$ ("$Ni^{III}$") was defined as the measurement for the pristine electrode at 9.5 keV while surface nickel in reduced state was referred to the lab-based XPS measurement at 1.5 keV ("$Ni^{II}$"). To define a lineshape for bulk charged nickel in $NiO_2$, the spectra for $Li_{0.01}NiO_2$ measured at 9.5 keV was decomposed including the $Ni^{II}$ lineshape in the peak model while removing satellite peaks in the 860 - 863 eV range. This approximation is expected to overestimate the percentage of $Ni^{II}$ and therefore the surface layer thickness, presenting an upper limit for its estimate.

## Computational methods

The DFT calculations were carried out with the Wien2k code, using the Perdew-Burke-Ernzerhof generalized gradient approximation (PBE-GGA) exchange-correlation potential. The paramagnetic ground state electronic density was obtained with a 10x10x10 wavevector-grid in the full primitive Brillouin zone, a plane wave cutoff of $R_{MT}^{min}K_{max} = 7$ (with $R_{MT}^{min}$ the smallest atomic sphere radius and $K_{max}$ the largest k-vector), and convergence criteria of 1.36 meV/f.u. for the total energy and $10^{-3}$ electrons/f.u. for the charge, respectively. The $LiNiO_2$ and $NiO_2$ crystal structures were derived from the Rietveld refinements of XRD data for pristine and deeply charged samples, respectively (**Supplementary Table 1**). All occupations of Li and Ni sites were approximated to unitary values. The muffin-tin radii for Li, Ni, and O were 1.69, 1.9, and 1.63

Bohr, respectively. The NiO structure was taken from Ref. [81] (Fm-3m, a = 4.177 Å). The muffin-tin radii for Ni and O in NiO were 2.11 and 1.81, respectively.

The converged electronic structures were used to build the low energy tight binding model by the wannierization routine of Wien2wannier and Wannier90 including all Ni 3d and O 2p bands (energy window from -8 to 4 eV) and a local rotation towards the Ni-O bonds in the $NiO_6$ octahedra. Such low energy model was used to compute the screened coulomb interactions using the gap2c code based on constrained random phase approximation (cRPA) theory [97]. A 5x5x5 kmesh and upper-cut-off energy of 15 Ry were found sufficient to obtained converged static U and J values within less than 0.01 eV. Two models were considered by excluding both Ni 3d and O 2p states or only the Ni 3d states in the correlated subspace, herein named *dp-dp* and *d-dp* models, respectively (See **Supplementary Note 4** for further details).

Cluster model simulations of the Ni 2p spectra were performed via the exact diagonalization method employed in the Quanty code. The cluster model Hamiltonian is $H^{cluster} = H^{(1)} + H_{dd}^{U} + H_{d}^{SO} - H_{dc}^{AMF}$, where $H_{dd}^{U} = \sum_{k=0,2,4} F_{dd}^{k} H^{F^k}$ describes the Coulomb interactions for the Ni 3d shell, $H_{SO}^{d} = \xi_{Co3d} \sum_i l_i \cdot s_i$ is the spin-orbit coupling for the Ni 3d shell, $H_{dc}^{AMF}$ is the double counting correction with the around mean field approximation, and $H^{(1)} = \sum_i \varepsilon_i d_i^{\dagger} d_i + \sum_{i \neq j} t_{ij} d_i^{\dagger} d_j + \sum_i \varepsilon_i p_i^{\dagger} p_i + \sum_{i \neq j} t_{ij} p_i^{\dagger} p_j + \sum_{ij} t_{ij} \left( d_i^{\dagger} p_j + h.c. \right)$ is the single electron part containing hopping and onsite energies obtained by the tight binding Hamiltonian of wannier90 (more details on this procedure can be found in Ref. [52]).

The static interactions obtained by the cRPA calculations were used for the Ni 3d intrashell Coulomb interactions in the cluster calculations. The $F_{dd}^0$ Slater parameter was taken as average of the opposite-spin 2-index U matrix and the $F_{dd}^2$ and $F_{dd}^4$ parameters were calculated from the Hund's coupling using the common relations $J = \frac{(F_{dd}^2+F_{dd}^4)}{14}$ and $\frac{F_{dd}^4}{F_{dd}^2} = 0.63$. For the ground state calculation, we did not find necessary to correct the onsite energies via a configuration interaction model since the cRPA-derived parameters already account for the p-d charge transfer at the RPA level in the *d-dp* model by including p-d screening channels in the polarization. The charge transfer parameter was thus calculated as $\Delta = E[d^{n+1}] - E[d^n]$, with $E[d^{n+1}]$ and $E[d^n]$ the total energies for the constrained ground state calculation configurations with $n+1$ and $n$ electrons in the Ni 3d shell ($n$ being the nominal one). The validity of this approach was verified by comparing the result for this calculation with one for which Δ was fit to the experiment. Due to the complex spectra for the electrode samples, we used the simpler case of NiO as test. As shown in **Supplementary Supplementary Figure 11**, the results are highly comparable, implying that the cRPA parameter can account well for the p-d charge transfer in this case.

To simulate the Ni 2p XPS spectra, $H^{cluster}$ was augmented with the Ni 2p core level. The Ni 2p - Ni 3d multiplet interactions were taken into account by Slater parametrization using 70 % scaled Slater parameters obtained for the related isolated single ion (taken from ref [98]). However, the well-screened monopole interaction was instead set to the cRPA value by $F_{pd}^0 = U_{pd} + \frac{1}{15} G_{pd}^1 + \frac{3}{70} G_{pd}^3$ with $U_{pd} = 1.2 \cdot U_{dd}$ and $U_{dd} = F_{dd}^0 - \frac{2}{63}(F_{dd}^2 + F_{dd}^4)$. In this case, the onsite energies were corrected by the charge transfer parameter within the configuration interaction model by setting the trace of Ni 2p, Ni 3d and O 2p onsite energies to $\varepsilon_p = -nU_{pd}$, $\varepsilon_d = \frac{36\,\Delta - \mathrm{n}\,(n+71)\frac{U_{dd}}{2} - 216U_{pd}}{n+36}$, and $\varepsilon_L = \varepsilon_d + nU_{dd} + 6U_{pd} - \Delta$, respectively.

The photoemission spectra $I$ was calculated with the Green function method as $I = -\mathrm{Im}\left\langle \psi_{GS} \middle| T^\dagger \frac{1}{\omega - H + i\Gamma/2} T \middle| \psi_{GS} \right\rangle$, with $\psi_{GS}$ the ground state wave function, $H$ the Hamiltonian

described above, *T* the transition operator (annihilation of Ni 2p core level) and Γ= 0.1 eV the core level lifetime. For the partial spectra calculation, *T* was restricted to include only one of the possible $d^n$ configurations (e.g. 6<n<10 for $NiO_2$). The list of all parameters is given in **Table 1**.

# REFERENCES

(1) Bianchini, M.; Roca-Ayats, M.; Hartmann, P.; Brezesinski, T.; Janek, J. There and Back Again—The Journey of $LiNiO_2$ as a Cathode Active Material. *Angew. Chem. Int. Ed.* **2019**, *58* (31), 10434–10458. https://doi.org/10.1002/anie.201812472.

(2) Radin, M. D.; Hy, S.; Sina, M.; Fang, C.; Liu, H.; Vinckeviciute, J.; Zhang, M.; Whittingham, M. S.; Meng, Y. S.; Van der Ven, A. Narrowing the Gap between Theoretical and Practical Capacities in Li-Ion Layered Oxide Cathode Materials. *Adv. Energy Mater.* **2017**, *7* (20), 1602888. https://doi.org/10.1002/aenm.201602888.

(3) Noh, H.-J.; Youn, S.; Yoon, C. S.; Sun, Y.-K. Comparison of the Structural and Electrochemical Properties of Layered $Li[Ni_xCo_yMn_z]O_2$ (x = 1/3, 0.5, 0.6, 0.7, 0.8 and 0.85) Cathode Material for Lithium-Ion Batteries. *J. Power Sources* **2013**, *233*, 121–130. https://doi.org/10.1016/j.jpowsour.2013.01.063.

(4) Friedrich, F.; Strehle, B.; Freiberg, A. T. S.; Kleiner, K.; Day, S. J.; Erk, C.; Piana, M.; Gasteiger, H. A. Capacity Fading Mechanisms of NCM-811 Cathodes in Lithium-Ion Batteries Studied by X-Ray Diffraction and Other Diagnostics. *J. Electrochem. Soc.* **2019**, *166* (15), A3760–A3774. https://doi.org/10.1149/2.0821915jes.

(5) Lin, F.; Markus, I. M.; Nordlund, D.; Weng, T.-C.; Asta, M. D.; Xin, H. L.; Doeff, M. M. Surface Reconstruction and Chemical Evolution of Stoichiometric Layered Cathode Materials for Lithium-Ion Batteries. *Nat. Commun.* **2014**, *5* (1), 3529. https://doi.org/10.1038/ncomms4529.

(6) Lin, F.; Nordlund, D.; Markus, I. M.; Weng, T.-C.; Xin, H. L.; Doeff, M. M. Profiling the Nanoscale Gradient in Stoichiometric Layered Cathode Particles for Lithium-Ion Batteries. *Energy Environ. Sci.* **2014**, *7* (9), 3077. https://doi.org/10.1039/C4EE01400F.

(7) Gauthier, M.; Carney, T. J.; Grimaud, A.; Giordano, L.; Pour, N.; Chang, H.-H.; Fenning, D. P.; Lux, S. F.; Paschos, O.; Bauer, C.; Maglia, F.; Lupart, S.; Lamp, P.; Shao-Horn, Y. Electrode–Electrolyte Interface in Li-Ion Batteries: Current Understanding and New Insights. *J. Phys. Chem. Lett.* **2015**, *6* (22), 4653–4672. https://doi.org/10.1021/acs.jpclett.5b01727.

(8) Xu, C.; Märker, K.; Lee, J.; Mahadevegowda, A.; Reeves, P. J.; Day, S. J.; Groh, M. F.; Emge, S. P.; Ducati, C.; Layla Mehdi, B.; Tang, C. C.; Grey, C. P. Bulk Fatigue Induced by Surface Reconstruction in Layered Ni-Rich Cathodes for Li-Ion Batteries. *Nat. Mater.* **2021**, *20* (1), 84–92. https://doi.org/10.1038/s41563-020-0767-8.

(9) Bautista Quisbert, E.; Fauth, F.; Abakumov, A. M.; Blangero, M.; Guignard, M.; Delmas, C. Understanding the High Voltage Behavior of $LiNiO_2$ Through the Electrochemical Properties of the Surface Layer. *Small* **2023**, 2300616. https://doi.org/10.1002/smll.202300616.

(10) Jung, R.; Metzger, M.; Maglia, F.; Stinner, C.; Gasteiger, H. A. Oxygen Release and Its Effect on the Cycling Stability of $LiNi_xMn_yCo_zO_2$ (NMC) Cathode Materials for Li-Ion Batteries. *J. Electrochem. Soc.* **2017**, *164* (7), A1361–A1377. https://doi.org/10.1149/2.0021707jes.

(11) Kaufman, L. A.; Huang, T.-Y.; Lee, D.; McCloskey, B. D. Particle Surface Cracking Is Correlated with Gas Evolution in High-Ni Li-Ion Cathode Materials. *ACS Appl. Mater. Interfaces* **2022**, *14* (35), 39959–39964. https://doi.org/10.1021/acsami.2c09194.

(12) Oswald, S.; Gasteiger, H. A. The Structural Stability Limit of Layered Lithium Transition Metal Oxides Due to Oxygen Release at High State of Charge and Its Dependence on the Nickel Content. *J. Electrochem. Soc.* **2023**, *170* (3), 030506. https://doi.org/10.1149/1945-7111/acbf80.

(13) Riewald, F.; Kurzhals, P.; Bianchini, M.; Sommer, H.; Janek, J.; Gasteiger, H. A. The $LiNiO_2$ Cathode Active Material: A Comprehensive Study of Calcination Conditions and Their Correlation with Physicochemical Properties Part II. Morphology. *J. Electrochem. Soc.* **2022**, *169* (2), 020529. https://doi.org/10.1149/1945-7111/ac4bf3.

(14) Pan, R.; Jo, E.; Cui, Z.; Manthiram, A. Degradation Pathways of Cobalt-Free $LiNiO_2$ Cathode in Lithium Batteries. *Adv. Funct. Mater.* **2023**, *33* (10), 2211461. https://doi.org/10.1002/adfm.202211461.

(15) Kondrakov, A. O.; Schmidt, A.; Xu, J.; Geßwein, H.; Mönig, R.; Hartmann, P.; Sommer, H.; Brezesinski, T.; Janek, J. Anisotropic Lattice Strain and Mechanical Degradation of High- and Low-Nickel NCM Cathode Materials for Li-Ion Batteries. *J. Phys. Chem. C* **2017**, *121* (6), 3286–3294. https://doi.org/10.1021/acs.jpcc.6b12885.

(16) Manthiram, A. A Reflection on Lithium-Ion Battery Cathode Chemistry. *Nat. Commun.* **2020**, *11* (1), 1550. https://doi.org/10.1038/s41467-020-15355-0.

(17) Rougier, A.; Delmas, C.; Chadwick, A. V. Non-Cooperative Jahn-Teller Effect in $LiNiO_2$: And EXAFS Study. *Solid State Commun.* **1995**, *94* (2), 123–127.

(18) Chen, H.; Freeman, C. L.; Harding, J. H. Charge Disproportionation and Jahn-Teller Distortion in $LiNiO_2$ and $NaNiO_2$: A Density Functional Theory Study. *Phys. Rev. B* **2011**, *84* (8), 085108. https://doi.org/10.1103/PhysRevB.84.085108.

(19) Nakai, I.; Takahashi, K.; Shiraishi, Y.; Nakagome, T.; Nishikawa, F. Study of the Jahn–Teller Distortion in $LiNiO_2$, a Cathode Material in a Rechargeable Lithium Battery, by in Situ X-Ray Absorption Fine Structure Analysis. *J. Solid State Chem.* **1998**, *140* (1), 145–148. https://doi.org/10.1006/jssc.1998.7943.

(20) Chung, J.-H.; Proffen, Th.; Shamoto, S.; Ghorayeb, A. M.; Croguennec, L.; Tian, W.; Sales, B. C.; Jin, R.; Mandrus, D.; Egami, T. Local Structure of $LiNiO_2$ Studied by Neutron Diffraction. *Phys. Rev. B* **2005**, *71* (6), 064410. https://doi.org/10.1103/PhysRevB.71.064410.

(21) van Elp, J.; Eskes, H.; Kuiper, P.; Sawatzky, G. A. Electronic Structure of Li-Doped NiO. *Phys. Rev. B* **1992**, *45* (4), 1612–1622. https://doi.org/10.1103/PhysRevB.45.1612.

(22) Kuiper, P.; Kruizinga, G.; Ghijsen, J.; Sawatzky, G. A.; Verweij, H. Character of Holes in $Li_xNi_{1-x}O$ and Their Magnetic Behavior. *Phys. Rev. Lett.* **1989**, *62* (2), 221–224. https://doi.org/10.1103/PhysRevLett.62.221.

(23) Kumara, L. S. R.; Sakata, O.; Yang, A.; Yamauchi, R.; Taguchi, M.; Matsuda, A.; Yoshimoto, M. Hard X-Ray Photoelectron Spectroscopy of $Li_xNi_{1-x}O$ Epitaxial Thin Films with a High Lithium Content. *J. Chem. Phys.* **2014**, *141* (4), 044718. https://doi.org/10.1063/1.4891366.

(24) Bisogni, V.; Catalano, S.; Green, R. J.; Gibert, M.; Scherwitzl, R.; Huang, Y.; Strocov, V. N.; Zubko, P.; Balandeh, S.; Triscone, J.-M.; Sawatzky, G.; Schmitt, T. Ground-State Oxygen Holes and the Metal–Insulator Transition in the Negative Charge-Transfer Rare-Earth Nickelates. *Nat. Commun.* **2016**, *7* (1), 13017. https://doi.org/10.1038/ncomms13017.

(25) Green, R. J.; Haverkort, M. W.; Sawatzky, G. A. Bond Disproportionation and Dynamical Charge Fluctuations in the Perovskite Rare-Earth Nickelates. *Phys. Rev. B* **2016**, *94* (19), 195127. https://doi.org/10.1103/PhysRevB.94.195127.

(26) Sawatzky, G. A.; Green, R. J. G. Sawatzky and R. J. Green, The Explicit Role of Anion States in High-Valence Metal Oxides, in Quantum Materials: Experiments and Theory (Forschungszentrum, Zentralbibliothek, Jülich, 2016). In *Quantum Materials: Experiments and Theory*; Forschungszentrum, Zentralbibliothek,: Jülich, 2016.

(27) Foyevtsova, K.; Elfimov, I.; Rottler, J.; Sawatzky, G. A. $LiNiO_2$ as a High-Entropy Charge- and Bond-Disproportionated Glass. *Phys. Rev. B* **2019**, *100* (16), 165104. https://doi.org/10.1103/PhysRevB.100.165104.

(28) Korotin, Dm. M.; Novoselov, D.; Anisimov, V. I. Paraorbital Ground State of the Trivalent Ni Ion in $LiNiO_2$ from DFT+DMFT Calculations. *Phys. Rev. B* **2019**, *99* (4), 045106. https://doi.org/10.1103/PhysRevB.99.045106.

(29) Genreith-Schriever, A. R.; Banerjee, H.; Menon, A. S.; Bassey, E. N.; Piper, L. F. J.; Grey, C. P.; Morris, A. J. Oxygen Hole Formation Controls Stability in $LiNiO_2$ Cathodes. *Joule* **2023**, *7* (7), 1623–1640. https://doi.org/10.1016/j.joule.2023.06.017.

(30) Huang, H.; Chang, Y.-C.; Huang, Y.-C.; Li, L.; Komarek, A. C.; Tjeng, L. H.; Orikasa, Y.; Pao, C.-W.; Chan, T.-S.; Chen, J.-M.; Haw, S.-C.; Zhou, J.; Wang, Y.; Lin, H.-J.; Chen, C.-T.; Dong, C.-L.; Kuo, C.-Y.; Wang, J.-Q.; Hu, Z.; Zhang, L. Unusual Double Ligand Holes as Catalytic Active Sites in LiNiO2. *Nat. Commun.* **2023**, *14* (1), 2112. https://doi.org/10.1038/s41467-023-37775-4.

(31) Zhang, M.; Kitchaev, D. A.; Lebens-Higgins, Z.; Vinckeviciute, J.; Zuba, M.; Reeves, P. J.; Grey, C. P.; Whittingham, M. S.; Piper, L. F. J.; Van der Ven, A.; Meng, Y. S. Pushing the Limit of 3d Transition Metal-Based Layered Oxides That Use Both Cation and Anion Redox for Energy Storage. *Nat. Rev. Mater.* **2022**, *7* (7), 522–540. https://doi.org/10.1038/s41578-022-00416-1.

(32) Menon, A. S.; Johnston, B. J.; Booth, S. G.; Zhang, L.; Kress, K.; Murdock, B. E.; Paez Fajardo, G.; Anthonisamy, N. N.; Tapia-Ruiz, N.; Agrestini, S.; Garcia-Fernandez, M.; Zhou, K.; Thakur, P. K.; Lee, T. L.; Nedoma, A. J.; Cussen, S. A.; Piper, L. F. J. Oxygen-Redox Activity in Non-Lithium-Excess Tungsten-Doped $LiNiO_2$ Cathode. *PRX Energy* **2023**, *2* (1), 013005. https://doi.org/10.1103/PRXEnergy.2.013005.

(33) Li, N.; Sallis, S.; Papp, J. K.; Wei, J.; McCloskey, B. D.; Yang, W.; Tong, W. Unraveling the Cationic and Anionic Redox Reactions in a Conventional Layered Oxide Cathode. *ACS Energy Lett.* **2019**, *4* (12), 2836–2842. https://doi.org/10.1021/acsenergylett.9b02147.

(34) Gent, W. E.; Lim, K.; Liang, Y.; Li, Q.; Barnes, T.; Ahn, S.-J.; Stone, K. H.; McIntire, M.; Hong, J.; Song, J. H.; Li, Y.; Mehta, A.; Ermon, S.; Tyliszczak, T.; Kilcoyne, D.; Vine, D.; Park, J.-H.; Doo, S.-K.; Toney, M. F.; Yang, W.; Prendergast, D.; Chueh, W. C. Coupling between Oxygen Redox and Cation Migration Explains Unusual Electrochemistry in Lithium-Rich Layered Oxides. *Nat. Commun.* **2017**, *8* (1), 2091. https://doi.org/10.1038/s41467-017-02041-x.

(35) Yang, W.; Devereaux, T. P. Anionic and Cationic Redox and Interfaces in Batteries: Advances from Soft X-Ray Absorption Spectroscopy to Resonant Inelastic Scattering. *J. Power Sources* **2018**, *389*, 188–197. https://doi.org/10.1016/j.jpowsour.2018.04.018.

(36) Luo, K.; Roberts, M. R.; Guerrini, N.; Tapia-Ruiz, N.; Hao, R.; Massel, F.; Pickup, D. M.; Ramos, S.; Liu, Y.-S.; Guo, J.; Chadwick, A. V.; Duda, L. C.; Bruce, P. G. Anion Redox Chemistry in the Cobalt Free 3d Transition Metal Oxide Intercalation Electrode $Li[Li_{0.2}Ni_{0.2}Mn_{0.6}]O_2$. *J. Am. Chem. Soc.* **2016**, *138* (35), 11211–11218. https://doi.org/10.1021/jacs.6b05111.

(37) Dai, K.; Wu, J.; Zhuo, Z.; Li, Q.; Sallis, S.; Mao, J.; Ai, G.; Sun, C.; Li, Z.; Gent, W. E.; Chueh, W. C.; Chuang, Y.; Zeng, R.; Shen, Z.; Pan, F.; Yan, S.; Piper, L. F. J.; Hussain, Z.; Liu, G.; Yang, W. High Reversibility of Lattice Oxygen Redox Quantified by Direct Bulk

Probes of Both Anionic and Cationic Redox Reactions. *Joule* **2019**, *3* (2), 518–541. https://doi.org/10.1016/j.joule.2018.11.014.

(38) Assat, G.; Tarascon, J.-M. Fundamental Understanding and Practical Challenges of Anionic Redox Activity in Li-Ion Batteries. *Nat. Energy* **2018**, *3* (5), 373–386. https://doi.org/10.1038/s41560-018-0097-0.

(39) Lee, G.; Wu, J.; Kim, D.; Cho, K.; Cho, M.; Yang, W.; Kang, Y. Reversible Anionic Redox Activities in Conventional $LiNi_{1/3}Co_{1/3}Mn_{1/3}O_2$ Cathodes. *Angew. Chem. Int. Ed.* **2020**, *59* (22), 8681–8688. https://doi.org/10.1002/anie.202001349.

(40) Kleiner, K.; Murray, C. A.; Grosu, C.; Ying, B.; Winter, M.; Nagel, P.; Schuppler, S.; Merz, M. On the Origin of Reversible and Irreversible Reactions in $LiNi_xCo_{(1-x)/2}Mn_{(1-x)/2}O_2$. *J. Electrochem. Soc.* **2021**, *168* (12), 120533. https://doi.org/10.1149/1945-7111/ac3c21.

(41) Yoon, W.-S.; Balasubramanian, M.; Chung, K. Y.; Yang, X.-Q.; McBreen, J.; Grey, C. P.; Fischer, D. A. Investigation of the Charge Compensation Mechanism on the Electrochemically Li-Ion Deintercalated $Li_{1-x}Co_{1/3}Ni_{1/3}Mn_{1/3}O_2$ Electrode System by Combination of Soft and Hard X-Ray Absorption Spectroscopy. *J. Am. Chem. Soc.* **2005**, *127* (49), 17479–17487. https://doi.org/10.1021/ja0530568.

(42) Kondrakov, A. O.; Geßwein, H.; Galdina, K.; de Biasi, L.; Meded, V.; Filatova, E. O.; Schumacher, G.; Wenzel, W.; Hartmann, P.; Brezesinski, T.; Janek, J. Charge-Transfer-Induced Lattice Collapse in Ni-Rich NCM Cathode Materials during Delithiation. *J. Phys. Chem. C* **2017**, *121* (44), 24381–24388. https://doi.org/10.1021/acs.jpcc.7b06598.

(43) Philippe, B.; Hahlin, M.; Edström, K.; Gustafsson, T.; Siegbahn, H.; Rensmo, H. Photoelectron Spectroscopy for Lithium Battery Interface Studies. *J. Electrochem. Soc.* **2016**, *163* (2), A178–A191. https://doi.org/10.1149/2.0051602jes.

(44) Källquist, I.; Le Ruyet, R.; Liu, H.; Mogensen, R.; Lee, M.-T.; Edström, K.; Naylor, A. J. Advances in Studying Interfacial Reactions in Rechargeable Batteries by Photoelectron Spectroscopy. *J. Mater. Chem. A* **2022**, *10* (37), 19466–19505. https://doi.org/10.1039/D2TA03242B.

(45) Laurita, A.; Zhu, L.; Cabelguen, P.-E.; Auvergniot, J.; Hamon, J.; Guyomard, D.; Dupré, N.; Moreau, P. Pristine Surface of Ni-Rich Layered Transition Metal Oxides as a Premise of Surface Reactivity. *ACS Appl. Mater. Interfaces* **2022**, *14* (37), 41945–41956. https://doi.org/10.1021/acsami.2c09358.

(46) Hartmann, L.; Pritzl, D.; Beyer, H.; Gasteiger, H. A. Evidence for $Li^+$/$H^+$ Exchange during Ambient Storage of Ni-Rich Cathode Active Materials. *J. Electrochem. Soc.* **2021**, *168* (7), 070507. https://doi.org/10.1149/1945-7111/ac0d3a.

(47) Lee, W.; Lee, S.; Lee, E.; Choi, M.; Thangavel, R.; Lee, Y.; Yoon, W.-S. Destabilization of the Surface Structure of Ni-Rich Layered Materials by Water-Washing Process. *Energy Storage Mater.* **2022**, *44*, 441–451. https://doi.org/10.1016/j.ensm.2021.11.006.

(48) Cherkashinin, G.; Motzko, M.; Schulz, N.; Späth, T.; Jaegermann, W. Electron Spectroscopy Study of Li[Ni,Co,Mn]$O_2$/Electrolyte Interface: Electronic Structure, Interface Composition, and Device Implications. *Chem. Mater.* **2015**, *27* (8), 2875–2887. https://doi.org/10.1021/cm5047534.

(49) Laubach, S.; Laubach, S.; Schmidt, P. C.; Ensling, D.; Schmid, S.; Jaegermann, W.; Thißen, A.; Nikolowski, K.; Ehrenberg, H. Changes in the Crystal and Electronic Structure of $LiCoO_2$ and $LiNiO_2$ upon Li Intercalation and de-Intercalation. *Phys. Chem. Chem. Phys.* **2009**, *11* (17), 3278. https://doi.org/10.1039/b901200a.

(50) Dahéron, L.; Dedryvère, R.; Martinez, H.; Ménétrier, M.; Denage, C.; Delmas, C.; Gonbeau, D. Electron Transfer Mechanisms upon Lithium Deintercalation from $LiCoO_2$ to $CoO_2$ Investigated by XPS. *Chem. Mater.* **2008**, *20* (2), 583–590. https://doi.org/10.1021/cm702546s.

(51) Ensling, D.; Cherkashinin, G.; Schmid, S.; Bhuvaneswari, S.; Thissen, A.; Jaegermann, W. Nonrigid Band Behavior of the Electronic Structure of $LiCoO_2$ Thin Film during Electrochemical Li Deintercalation. *Chem. Mater.* **2014**, *26* (13), 3948–3956. https://doi.org/10.1021/cm501480b.

(52) Fantin, R.; van Roekeghem, A.; Benayad, A. Self-Regulated Ligand-Metal Charge Transfer upon Lithium-Ion Deintercalation Process from $LiCoO_2$ to $CoO_2$. *PRX Energy* **2023**, *2* (4), 043010. https://doi.org/10.1103/PRXEnergy.2.043010.

(53) Assat, G.; Iadecola, A.; Foix, D.; Dedryvère, R.; Tarascon, J.-M. Direct Quantification of Anionic Redox over Long Cycling of Li-Rich NMC via Hard X-Ray Photoemission Spectroscopy. *ACS Energy Lett.* **2018**, *3* (11), 2721–2728. https://doi.org/10.1021/acsenergylett.8b01798.

(54) Li, B.; Kumar, K.; Roy, I.; Morozov, A. V.; Emelyanova, O. V.; Zhang, L.; Koç, T.; Belin, S.; Cabana, J.; Dedryvère, R.; Abakumov, A. M.; Tarascon, J.-M. Capturing Dynamic Ligand-to-Metal Charge Transfer with a Long-Lived Cationic Intermediate for Anionic Redox. *Nat. Mater.* **2022**, *21* (10), 1165–1174. https://doi.org/10.1038/s41563-022-01278-2.

(55) Naylor, A. J.; Makkos, E.; Maibach, J.; Guerrini, N.; Sobkowiak, A.; Björklund, E.; Lozano, J. G.; Menon, A. S.; Younesi, R.; Roberts, M. R.; Edström, K.; Islam, M. S.; Bruce, P. G. Depth-Dependent Oxygen Redox Activity in Lithium-Rich Layered Oxide Cathodes. *J. Mater. Chem. A* **2019**, *7* (44), 25355–25368. https://doi.org/10.1039/C9TA09019C.

(56) Lebens-Higgins, Z. W.; Chung, H.; Zuba, M. J.; Rana, J.; Li, Y.; Faenza, N. V.; Pereira, N.; McCloskey, B. D.; Rodolakis, F.; Yang, W.; Whittingham, M. S.; Amatucci, G. G.; Meng, Y. S.; Lee, T.-L.; Piper, L. F. J. How Bulk Sensitive Is Hard X-Ray Photoelectron

Spectroscopy: Accounting for the Cathode–Electrolyte Interface When Addressing Oxygen Redox. *J. Phys. Chem. Lett.* **2020**, *11* (6), 2106–2112. https://doi.org/10.1021/acs.jpclett.0c00229.

(57) Fantin, R.; Van Roekeghem, A.; Benayad, A. Revisiting Co 2p Core-level Photoemission in $LiCoO_2$ by In-lab Soft and Hard X-ray Photoelectron Spectroscopy: A Depth-dependent Study of Cobalt Electronic Structure. *Surf. Interface Anal.* **2022**, sia.7167. https://doi.org/10.1002/sia.7167.

(58) Renault, O.; Deleuze, P.-M.; Courtin, J.; Bure, T. R.; Gauthier, N.; Nolot, E.; Robert-Goumet, C.; Pauly, N.; Martinez, E.; Artyushkova, K. New Directions in the Analysis of Buried Interfaces for Device Technology by Hard X-Ray Photoemission. *Faraday Discuss.* **2022**, *236*, 288–310. https://doi.org/10.1039/D1FD00110H.

(59) Renault, O.; Martinez, E.; Zborowski, C.; Mann, J.; Inoue, R.; Newman, J.; Watanabe, K. Analysis of Buried Interfaces in Multilayer Device Structures with Hard XPS (HAXPES) Using a CrKα Source. *Surf. Interface Anal.* **2018**, *50* (11), 1158–1162. https://doi.org/10.1002/sia.6451.

(60) Hufner, S. *Photoelectron Spectroscopy*, Third Edition.; Springer, 2003.

(61) Gupta, A.; Chemelewski, W. D.; Buddie Mullins, C.; Goodenough, J. B. High-Rate Oxygen Evolution Reaction on Al-Doped $LiNiO_2$. *Adv. Mater.* **2015**, *27* (39), 6063–6067. https://doi.org/10.1002/adma.201502256.

(62) Zheng, X.; Li, X.; Wang, Z.; Guo, H.; Huang, Z.; Yan, G.; Wang, D. Investigation and Improvement on the Electrochemical Performance and Storage Characteristics of $LiNiO_2$-Based Materials for Lithium Ion Battery. *Electrochimica Acta* **2016**, *191*, 832–840. https://doi.org/10.1016/j.electacta.2016.01.142.

(63) Ren, X.; Wei, C.; Sun, Y.; Liu, X.; Meng, F.; Meng, X.; Sun, S.; Xi, S.; Du, Y.; Bi, Z.; Shang, G.; Fisher, A. C.; Gu, L.; Xu, Z. J. Constructing an Adaptive Heterojunction as a Highly Active Catalyst for the Oxygen Evolution Reaction. *Adv. Mater.* **2020**, *32* (30), 2001292. https://doi.org/10.1002/adma.202001292.

(64) Heo, K.; Lee, J.; Song, Y.-W.; Kim, M.-Y.; Jeong, H.; DoCheon, A.; Jaekook, K.; Lim, J. Synthesis and Electrochemical Performance Analysis of $LiNiO_2$ Cathode Material Using Taylor-Couette Flow-Type Co-Precipitation Method. *J. Electrochem. Soc.* **2021**, *168* (1), 010521. https://doi.org/10.1149/1945-7111/abd91a.

(65) Phan Nguyen, T.; Thi Giang, T.; Tae Kim, I. Restructuring NiO to $LiNiO_2$: Ultrastable and Reversible Anodes for Lithium-Ion Batteries. *Chem. Eng. J.* **2022**, *437*, 135292. https://doi.org/10.1016/j.cej.2022.135292.

(66) Thi Bich Tran, T.; Park, E.-J.; Kim, H.-I.; Lee, S.-H.; Jang, H.-J.; Son, J.-T. High Rate Performance of Lithium-Ion Batteries with Co-Free $LiNiO_2$ Cathode. *Mater. Lett.* **2022**, *316*, 131810. https://doi.org/10.1016/j.matlet.2022.131810.

(67) Ding, G.; Yao, M.; Li, J.; Yang, T.; Zhang, Y.; Liu, K.; Huang, X.; Wu, Z.; Chen, J.; Wu, Z.; Du, J.; Rong, C.; Liu, Q.; Zhang, W.; Cheng, F. Molten Salt-Assisted Synthesis of Single-Crystalline Nonstoichiometric $Li_{1+x}Ni_{1-x}O_2$ with Improved Structural Stability. *Adv. Energy Mater.* **2023**, *13* (23), 2300407. https://doi.org/10.1002/aenm.202300407.

(68) Liu, Z.; Wu, J.; Zeng, J.; Li, F.; Peng, C.; Xue, D.; Zhu, M.; Liu, J. Co-Free Layered Oxide Cathode Material with Stable Anionic Redox Reaction for Sodium-Ion Batteries. *Adv. Energy Mater.* **2023**, 2301471. https://doi.org/10.1002/aenm.202301471.

(69) Azmi, R.; Trouillet, V.; Strafela, M.; Ulrich, S.; Ehrenberg, H.; Bruns, M. Surface Analytical Approaches to Reliably Characterize Lithium Ion Battery Electrodes. *Surf. Interface Anal.* **2018**, *50* (1), 43–51. https://doi.org/10.1002/sia.6330.

(70) Azmi, R.; Masoumi, M.; Ehrenberg, H.; Trouillet, V.; Bruns, M. Surface Analytical Characterization of $LiNi_{0.8-y}Mn_yCo_{0.2}O_2$ ($0 \leq y \leq 0.4$) Compounds for Lithium-Ion Battery Electrodes. *Surf. Interface Anal.* **2018**, *50* (11), 1132–1137. https://doi.org/10.1002/sia.6415.

(71) Bondarchuk, O.; LaGrow, A. P.; Kvasha, A.; Thieu, T.; Ayerbe, E.; Urdampilleta, I. On the X-Ray Photoelectron Spectroscopy Analysis of $LiNi_xMn_yCo_zO_2$ Material and Electrodes. *Appl. Surf. Sci.* **2021**, *535*, 147699. https://doi.org/10.1016/j.apsusc.2020.147699.

(72) Fu, Z.; Hu, J.; Hu, W.; Yang, S.; Luo, Y. Quantitative Analysis of $Ni^{2+}/Ni^{3+}$ in $Li[Ni_xMn_yCo_z]O_2$ Cathode Materials: Non-Linear Least-Squares Fitting of XPS Spectra. *Appl. Surf. Sci.* **2018**, *441*, 1048–1056. https://doi.org/10.1016/j.apsusc.2018.02.114.

(73) Mock, M.; Bianchini, M.; Fauth, F.; Albe, K.; Sicolo, S. Atomistic Understanding of the $LiNiO_2$–$NiO_2$ Phase Diagram from Experimentally Guided Lattice Models. *J. Mater. Chem. A* **2021**, *9* (26), 14928–14940. https://doi.org/10.1039/D1TA00563D.

(74) Fantin, R.; Van Roekeghem, A.; Rueff, J.-P.; Benayad, A. Surface Analysis Insight Note: Accounting for X-Ray Beam Damage Effects in Positive Electrode-Electrolyte Interphase Investigations (under Submission).

(75) Marchesini, S.; Reed, B. P.; Jones, H.; Matjacic, L.; Rosser, T. E.; Zhou, Y.; Brennan, B.; Tiddia, M.; Jervis, R.; Loveridge, Melanie. J.; Raccichini, R.; Park, J.; Wain, A. J.; Hinds, G.; Gilmore, I. S.; Shard, A. G.; Pollard, A. J. Surface Analysis of Pristine and Cycled NMC/Graphite Lithium-Ion Battery Electrodes: Addressing the Measurement Challenges. *ACS Appl. Mater. Interfaces* **2022**, *14* (47), 52779–52793. https://doi.org/10.1021/acsami.2c13636.

(76) Vaugier, L.; Jiang, H.; Biermann, S. Hubbard U and Hund Exchange J in Transition Metal Oxides: Screening versus Localization Trends from Constrained Random Phase

Approximation. *Phys. Rev. B* **2012**, *86* (16), 165105. https://doi.org/10.1103/PhysRevB.86.165105.

(77) Biermann, S.; van Roekeghem, A. Electronic Polarons, Cumulants and Doubly Dynamical Mean Field Theory: Theoretical Spectroscopy for Correlated and Less Correlated Materials. *J. Electron Spectrosc. Relat. Phenom.* **2016**, *208*, 17–23. https://doi.org/10.1016/j.elspec.2016.01.001.

(78) Biesinger, M. C.; Payne, B. P.; Grosvenor, A. P.; Lau, L. W. M.; Gerson, A. R.; Smart, R. St. C. Resolving Surface Chemical States in XPS Analysis of First Row Transition Metals, Oxides and Hydroxides: Cr, Mn, Fe, Co and Ni. *Appl. Surf. Sci.* **2011**, *257* (7), 2717–2730. https://doi.org/10.1016/j.apsusc.2010.10.051.

(79) van Veenendaal, M. Competition between Screening Channels in Core-Level x-Ray Photoemission as a Probe of Changes in the Ground-State Properties of Transition-Metal Compounds. *Phys. Rev. B* **2006**, *74* (8), 085118. https://doi.org/10.1103/PhysRevB.74.085118.

(80) Ghiasi, M.; Hariki, A.; Winder, M.; Kuneš, J.; Regoutz, A.; Lee, T.-L.; Hu, Y.; Rueff, J.-P.; de Groot, F. M. F. Charge-Transfer Effect in Hard x-Ray 1 s and 2 p Photoemission Spectra: LDA + DMFT and Cluster-Model Analysis. *Phys. Rev. B* **2019**, *100* (7), 075146. https://doi.org/10.1103/PhysRevB.100.075146.

(81) Haverkort, M. W.; Zwierzycki, M.; Andersen, O. K. Multiplet Ligand-Field Theory Using Wannier Orbitals. *Phys. Rev. B* **2012**, *85* (16), 165113. https://doi.org/10.1103/PhysRevB.85.165113.

(82) Haverkort, M. W. *Quanty* for Core Level Spectroscopy - Excitons, Resonances and Band Excitations in Time and Frequency Domain. *J. Phys. Conf. Ser.* **2016**, *712*, 012001. https://doi.org/10.1088/1742-6596/712/1/012001.

(83) Bocquet, A. E.; Mizokawa, T.; Morikawa, K.; Fujimori, A.; Barman, S. R.; Maiti, K.; Sarma, D. D.; Tokura, Y.; Onoda, M. Electronic Structure of Early 3 *d* -Transition-Metal Oxides by Analysis of the 2 *p* Core-Level Photoemission Spectra. *Phys. Rev. B* **1996**, *53* (3), 1161–1170. https://doi.org/10.1103/PhysRevB.53.1161.

(84) Bocquet, A. E.; Mizokawa, T.; Saitoh, T.; Namatame, H.; Fujimori, A. Electronic Structure of 3 *d* -Transition-Metal Compounds by Analysis of the 2 *p* Core-Level Photoemission Spectra. *Phys. Rev. B* **1992**, *46* (7), 3771–3784. https://doi.org/10.1103/PhysRevB.46.3771.

(85) Zaanen, J.; Sawatzky, G. A.; Allen, J. W. Band Gaps and Electronic Structure of Transition-Metal Compounds. *Phys. Rev. Lett.* **1985**, *55* (4), 418–421. https://doi.org/10.1103/PhysRevLett.55.418.

(86) Taguchi, M.; Matsunami, M.; Ishida, Y.; Eguchi, R.; Chainani, A.; Takata, Y.; Yabashi, M.; Tamasaku, K.; Nishino, Y.; Ishikawa, T.; Senba, Y.; Ohashi, H.; Shin, S. Revisiting the Valence-Band and Core-Level Photoemission Spectra of NiO. *Phys. Rev. Lett.* **2008**, *100* (20), 206401. https://doi.org/10.1103/PhysRevLett.100.206401.

(87) Alders, D.; Voogt, F. C.; Hibma, T.; Sawatzky, G. A. Nonlocal Screening Effects in 2 *p* x-Ray Photoemission Spectroscopy of NiO (100). *Phys. Rev. B* **1996**, *54* (11), 7716–7719. https://doi.org/10.1103/PhysRevB.54.7716.

(88) Altieri, S.; Tjeng, L. H.; Tanaka, A.; Sawatzky, G. A. Core-Level x-Ray Photoemission on NiO in the Impurity Limit. *Phys. Rev. B* **2000**, *61* (20), 13403–13409. https://doi.org/10.1103/PhysRevB.61.13403.

(89) Flores, E.; Novák, P.; Berg, E. J. In Situ and Operando Raman Spectroscopy of Layered Transition Metal Oxides for Li-Ion Battery Cathodes. *Front. Energy Res.* **2018**, *6*, 82. https://doi.org/10.3389/fenrg.2018.00082.

(90) Flores, E.; Novák, P.; Aschauer, U.; Berg, E. J. Cation Ordering and Redox Chemistry of Layered Ni-Rich $Li_xNi_{1-2y}Co_yMn_yO_2$: An Operando Raman Spectroscopy Study. *Chem. Mater.* **2020**, *32* (1), 186–194. https://doi.org/10.1021/acs.chemmater.9b03202.

(91) Marianetti, C. A.; Morgan, D.; Ceder, G. First-Principles Investigation of the Cooperative Jahn-Teller Effect for Octahedrally Coordinated Transition-Metal Ions. *Phys. Rev. B* **2001**, *63* (22), 224304. https://doi.org/10.1103/PhysRevB.63.224304.

(92) Kieffer, J.; Valls, V.; Blanc, N.; Hennig, C. New Tools for Calibrating Diffraction Setups. *J. Synchrotron Radiat.* **2020**, *27* (2), 558–566. https://doi.org/10.1107/S1600577520000776.

(93) Rodríguez-Carvajal, J. Recent Advances in Magnetic Structure Determination by Neutron Powder Diffraction. *Phys. B Condens. Matter* **1993**, *192* (1–2), 55–69. https://doi.org/10.1016/0921-4526(93)90108-I.

(94) Razeg, K. H.; AL-Hilli, M. F.; Khalefa, A. A.; Aadim, K. A. Structural and Optical Properties of ($Li_xNi_{2-x}O_2$) Thin Films Deposited by Pulsed Laser Deposited (PLD) Technique at Different Doping Ratio. *Int. J. Phys.* **2017**, *5*, 46–52. https://doi.org/10.12691/ijp-5-2-3.

(95) El-Bana, M. S.; El Radaf, I. M.; Fouad, S. S.; Sakr, G. B. Structural and Optoelectrical Properties of Nanostructured $LiNiO_2$ Thin Films Grown by Spray Pyrolysis Technique. *J. Alloys Compd.* **2017**, *705*, 333–339. https://doi.org/10.1016/j.jallcom.2017.02.106.

(96) Tanuma, S.; Powell, C. J.; Penn, D. R. Calculations of Electron Inelastic Mean Free Paths. V. Data for 14 Organic Compounds over the 50-2000 EV Range. *Surf. Interface Anal.* **1994**, *21* (3), 165–176. https://doi.org/10.1002/sia.740210302.

(97) Jiang, H.; Blaha, P. G W with Linearized Augmented Plane Waves Extended by High-Energy Local Orbitals. *Phys. Rev. B* **2016**, *93* (11), 115203. https://doi.org/10.1103/PhysRevB.93.115203.

(98) Retegan, M. Crispy: V0.7.4, 2019. https://www.esrf.fr/computing/scientific/crispy/index.html.

## ACKNOWLEDGMENT

This work was supported by the "Recherches Technologiques de Base" program of the French National Research Agency (ANR) and by CEA FOCUS-Battery Program and was in part carried out at the platform of nano-characterization (PFNC) of CEA Grenoble. We acknowledge the Battery Interface Genome - Materials Acceleration Platform (Big-Map) EU project for providing the $LiNiO_2$ electrodes and materials for electrochemical characterization. The synchrotron HAXPES experiments were carried out within in-house Galaxies beam time at SOLEIL. Olivier Ulrich, Jean-Sébastien Micha, and Olivier Geaymond (BM32 beamline staff) are thanked for their help in the preparation of the synchrotron XRD experiments.

## AUTHOR CONTRIBUTIONS

R.F., A.V.R., and A.B. conceived the project and designed the experiments. R.F. prepared the ex-situ samples. R.F. performed the XPS and HAXPES experiments and data analysis under the supervision of A.B. R.F. performed the ab initio simulations under the supervision of A.V.R. T.J. performed the XRD experiments and data analysis. R.R. performed the Raman spectroscopy experiments and data analysis. G.L. performed the STEM-EELS experiments and data analysis. J.-P. R. performed the HAXPES synchrotron experiments and R.F. analyzed the data. R.F. and A.B. participated in the interpretation of all analyses. R.F. prepared the figures and wrote the manuscript with significant contributions from A.V.R and A.B. All authors participated in discussions, contributed to the writing of their respective experimental sections, and reviewed the manuscript as a whole.

# ADDITIONAL INFORMATION

**Supplementary Information**: electrochemical data, structural characterization including XRD, SEM, and EELS analysis, additional XPS data and analysis (IMFP calculation, pSEI thickness, quantification), calculated DFT electronic structures and details on the cRPA method.

# FIGURES

**Figure 1**

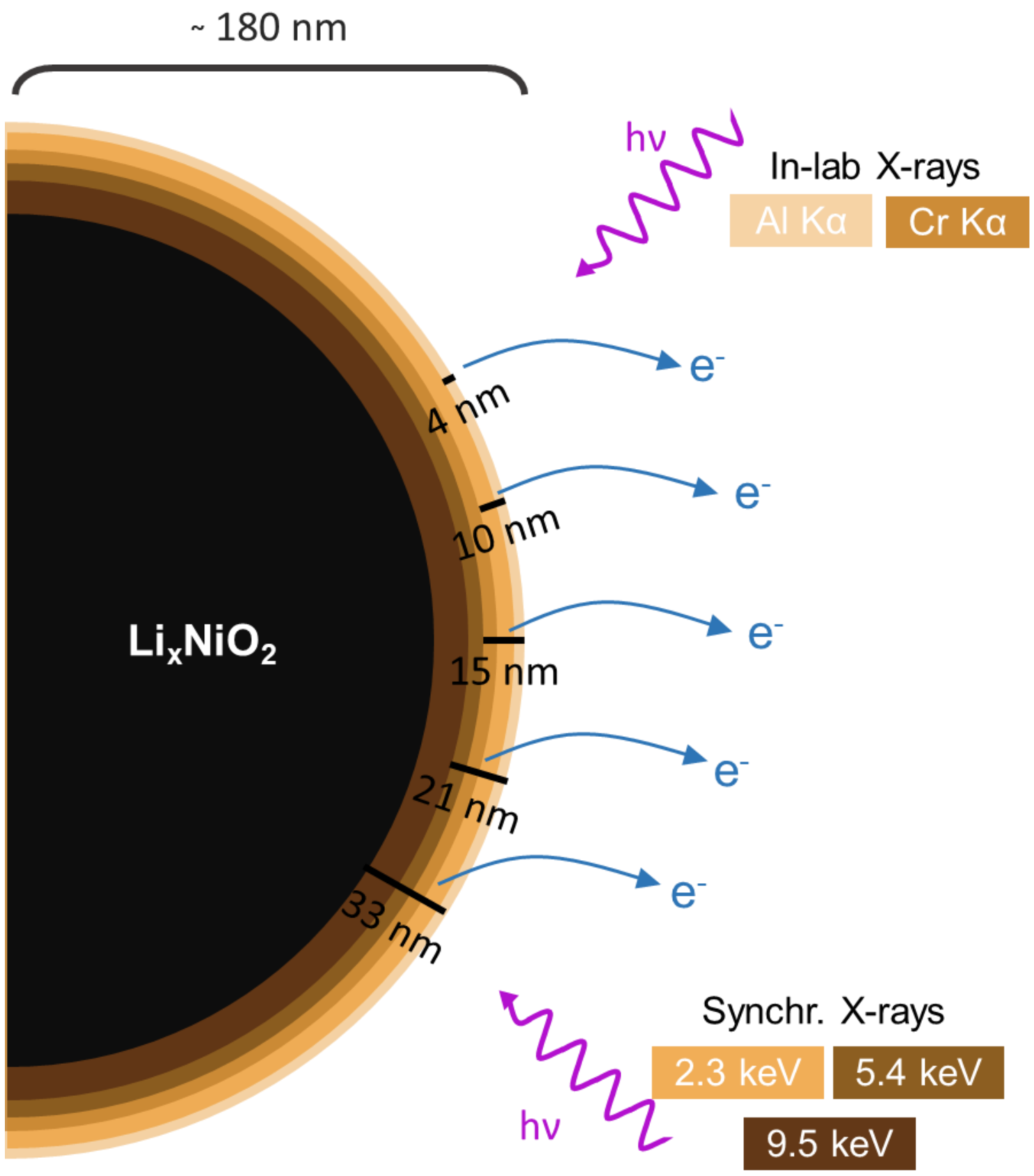

**Figure 1. Schematic illustration of the experimental approach of this study.** The figure is scaled to the average size of $Li_xNiO_2$ single particles estimated by SEM imaging (**Supplementary Fig. 3**) and the sampling depth depending on the X-ray photon energy and take-off angle of lab-based and synchrotron experiments. The inelastic mean free path of Ni 2p calculated using the TPP2M formula were used as representative (**Supplementary Tab. 2**).

**Figure 2**

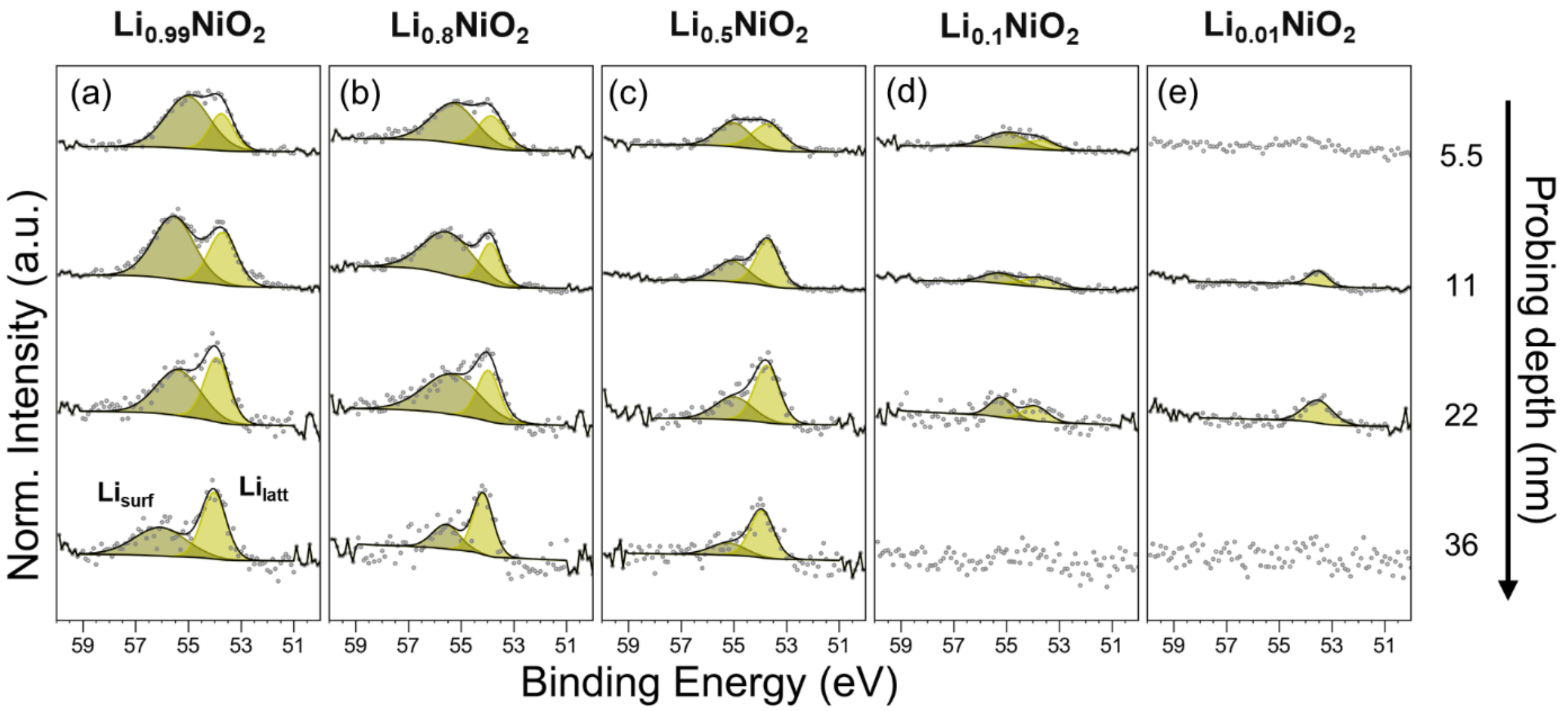

**Figure 2. Lithium non-destructive depth profile by energy-dependent HAXPES.** High-resolution Li 1s core-level spectra of (a) $LiNiO_2$, (b) $Li_{0.8}NiO_2$, (c) $Li_{0.5}NiO_2$, (d) $Li_{0.1}NiO_2$, and (e) $Li_{0.01}NiO_2$ electrodes. For each sample, the spectra are shown as stack plots from the most surface (top) to the most bulk-sensitive (bottom) measurement. The spectra were normalized to the average background intensity in the 48-50 eV region to allow peak intensity comparison. Experimental data, peak components and fits are shown as black dots, colored-filled areas and black lines, respectively.

**Figure 3**

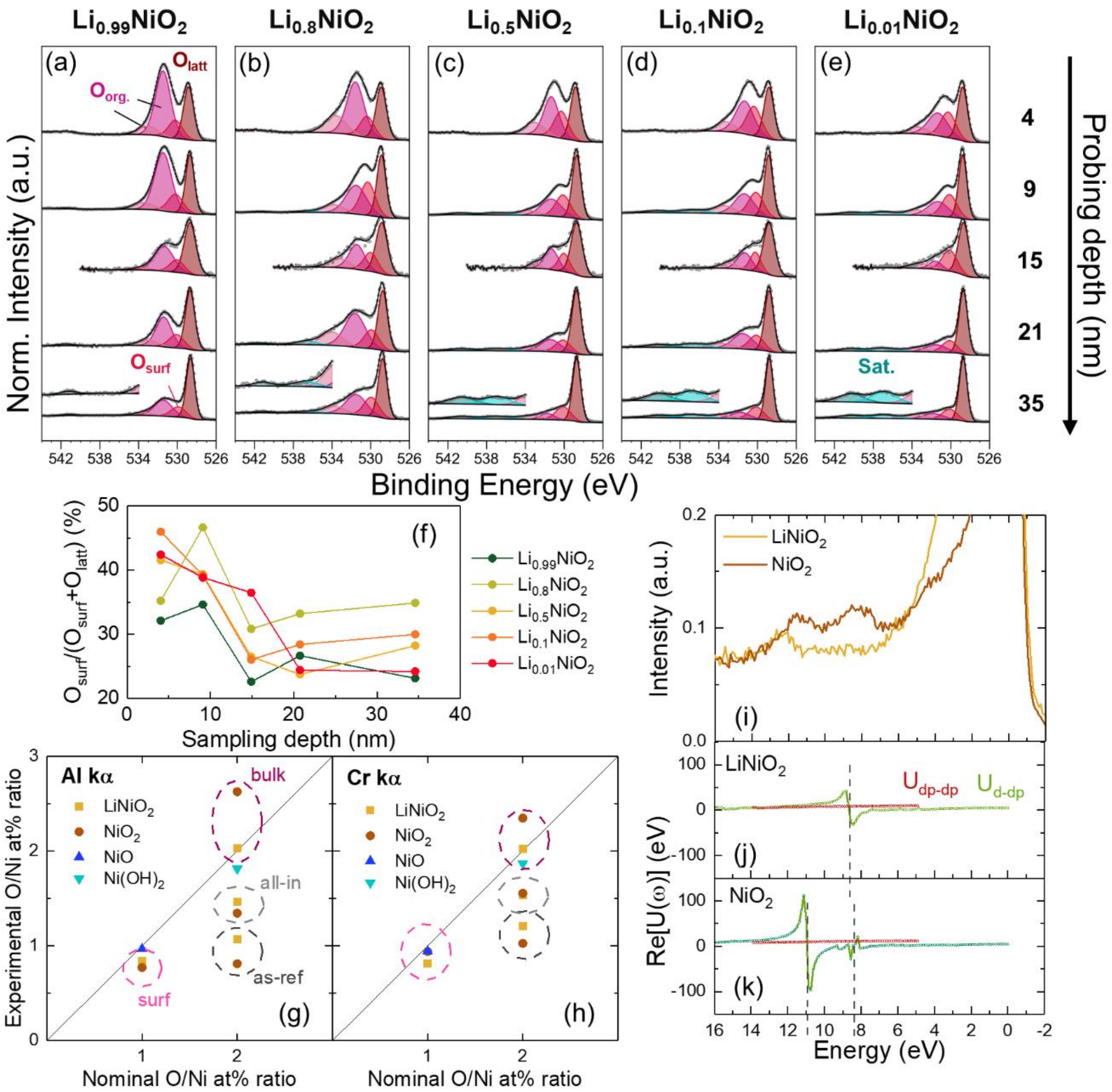

**Figure 3 Insights on oxygen surface-to-bulk evolution upon deintercalation.** (a-e) High-resolution O 1s spectra of the $Li_xNiO_2$ series of samples. All spectra were normalized to the lowest-binding energy peak intensity and are ordered from the most surface sensitive (top) to the most bulk sensitive (bottom). Experimental data, peak components and fits are shown as black dots, colored-filled areas and black lines, respectively. A magnification of the satellite region is shown for each spectrum acquired at 9.5 keV. (f) Calculated area percentage ratio of $O_{surf}$ to the sum of $O_{surf}$ and $O_{latt}$, as a function of the sampling depth for the O 1s photoemission of each XPS setting. Lab-based (g) XPS and (h) HAXPES quantification of the O/Ni at% ratio relative to nominal stoichiometry for reference NiO and $Ni(OH)_2$ samples as well as experimental $Li_{0.99}NiO_2$ and $Li_{0.01}NiO_2$. Comparison of (i) O 1s high-energy satellites of $LiNiO_2$ and $Li_{0.01}NiO_2$ with the screened Coulomb interaction calculated by cRPA for (j) $LiNiO_2$ and (k) $NiO_2$ structures using a d-dp (green) or dp-dp (red) model. Vertical dotted lines are shown to indicate the position of the plasmonic excitations.

**Figure 4**

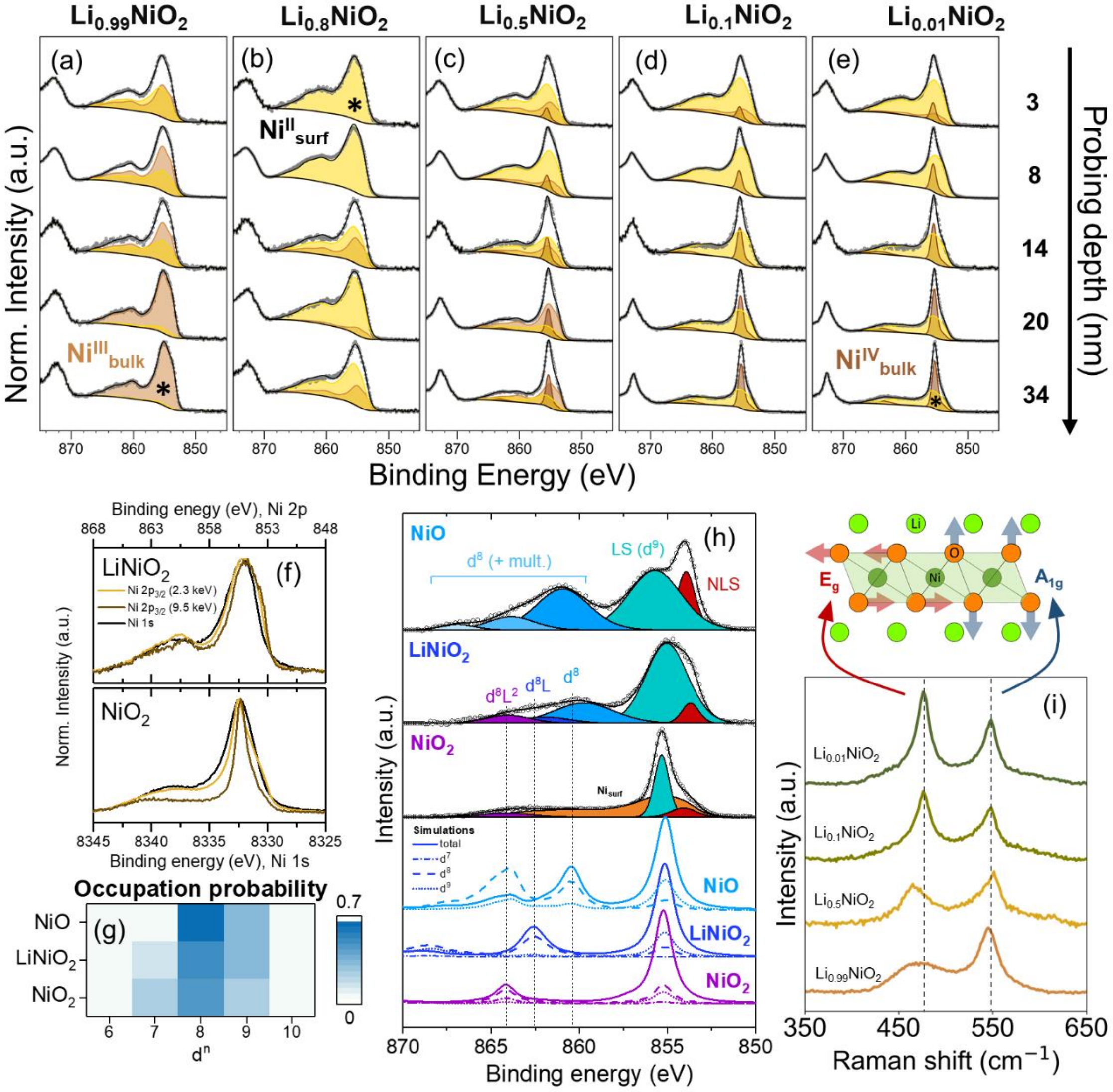

**Figure 4 Depth-dependent study of nickel electronic structure in $Li_xNiO_2$.** (a-e) High-resolution Ni $2p_{3/2}$ spectra of the $Li_xNiO_2$ series of samples. All spectra are normalized to the peak maximum and are stacked top-to-bottom in the order of increasing bulk sensitivity. Experimental data, peak components and fits are shown as black dots, colored-filled areas and black lines, respectively. The spectra used as line components are marked by a star symbol. (f) Comparison of Ni 1s (at 9.5 keV) and Ni $2p_{3/2}$ (at 2.3 and 9.5 keV) spectra for pristine $LiNiO_2$ (top) and deeply

deintercalated $Li_{0.01}NiO_2$ (bottom). To allow direct comparison, the spectra were normalized and shifted to the peak maximum. For reference, absolute binding energies for the Ni 1s and 2p spectra are shown in the bottom and top axes, respectively. (g) Ground state occupation probabilities of different Ni $3d^n$ configuration as resulted by cluster model calculations of NiO, $LiNiO_2$, and $NiO_2$. (h) Comparison of experimental and simulated Ni $2p_{3/2}$ spectra. The experimental data of $LiNiO_2$ and $NiO_2$ was peak fitted based on the NiO model, the surface contributions identified by the reference-based peak fitting approach, and the simulations for the bulk local electronic structure. For each simulated total spectra, partial contributions for $d^7$, $d^8$, and $d^9$ configurations are shown as dash-dot, dash and dot lines. These spectra were calculated by constraining the calculation to each specific configuration for both the initial and final state. (i) Evolution of local Ni-O bonding investigated by Raman spectroscopy. The dynamical effect of $E_g$ and $A_{1g}$ Raman modes is sketched above the spectra.

**Table 1**

| | $F^0_{dd}$ | $F^2_{dd}$ | $F^4_{dd}$ | $\Delta$ | $\xi_{3d}$ | $F^0_{pd}$ | $F^2_{pd}$ | $G^1_{pd}$ | $G^3_{pd}$ | $\xi_{2p}$ |
|---|---|---|---|---|---|---|---|---|---|---|
| **NiO** | 5.6 | 7.1 | 4.5 | 2.68 | 0.081 | 7.2 | 6.67 | 4.92 | 2.8 | 11.51 |
| **$LiNiO_2$** | 4.68 | 6.65 | 4.19 | -3.85 | 0.091 | 5.61 | 5.84 | 4.43 | 2.52 | 11.51 |
| **$NiO_2$** | 4.46 | 6.44 | 4.06 | -6.02 | 0.101 | 5.45 | 7.19 | 5.52 | 3.14 | 11.51 |

**Table 1. Parameters used for the cluster model simulations.**

# Supplementary information for

# Depth-resolving the redox compensation mechanism in $Li_xNiO_2$

Roberto Fantin,[a] Thibaut Jousseaume,[b] Raphael Ramos,[a] Gauthier Lefevre,[a] Ambroise Van Roekeghem,[a] Jean-Pascal Rueff,[c] Anass Benayad[a*]

[a] Univ. Grenoble Alpes, CEA-LITEN, Grenoble, 38054, France

[b] Univ. Grenoble Alpes, CEA-IRIG, Grenoble, 38054, Cedex 9, France

[c] Synchrotron SOLEIL, L'Orme des Merisiers, Saint-Aubin, BP 48, 91192 Gif-sur-Yvette Cedex, France

# Supplementary Notes

## Supplementary Note 1: XPS analysis of the CEI layer

The C 1s and F 1s spectra acquired with Al Kα are shown in **Supplementary Figure 4**. Because of the fast binder degradation measured during synchrotron experiments, the data will not be discussed in detail here but in a dedicated study [1]. Two peaks can be distinguished in the F 1s peak: the lower-energy one (~684 eV) was related to LiF while the other (~ 688 eV) to $CF_2$ component of PVDF. The presence of surface LiF is observed to decrease with increasing cut-off voltage of *ex-situ* samples. As can be seen, both peaks shift depending on the sample, indicating tendency of charging for these insulating species. In case of PVDF, it can depend on its local morphology in the area of analysis. For LiF instead, it can be related to the growth and composition of the CEI and buried NiO-like surface layer.

The C 1s XPS spectra is a convolution of many contributions, including surface species in $LiNiO_2$ particles, binder, and carbon black additive. Several constrains were therefore adopted. For the $CH_2$ and $CF_2$ components of PVDF, two GL(30) peak with energy separation of 4.5 eV and equal intensity and FHWM were used. A third higher energy peak separated by ~1.8 eV to $CF_2$ was also included, related to $CF_3$ environments. For carbon black, a single line-shape including asymmetric main line and π-π* satellite was obtained by fitting the reference material. Considering the O 1s spectra (see below), other components due to carbon bound to oxygen were also used.

To verify the peak assignation in O 1s spectra discussed in the main text, we relied on XPS quantification and reference samples. **Supplementary Figure 5** shows the comparison of the high resolution O 1s spectra for (a) pristine electrode, (b) bare $LiNiO_2$ powder, (c) PVDF reference powder, and (d) $Li_2CO_3$ acquired with the lab-based Al Kα source. The presence of C-O peak at 533 eV in the pristine PVDF powder suggests that a similar component should be presence in the O 1s spectra of the electrode. The peak at 531.5 was related instead to $Li_2CO_3$, although a shift is observed with respect to the pure material.

However, the quantitative XPS analysis for the pristine $LiNiO_2$ electrode and powder (**Supplementary Table 4** and**Supplementary** Table 5**, respectively)** suggest that assigning the peak to $Li_2CO_3$ only in O 1s spectra of the electrode sample is not consistent with the much lower amount of carbon. To explain this, we propose that LiOH is also present, since it has O 1s and Li 1s position close to $Li_2CO_3$ and its inclusion describes successfully the intensity of the O 1s and Li 1s spectra. In fact, by summing the contributions of F 1s and C1s related to LiF and $Li_2CO_3$, respectively, about 5 at% of the $Li_{CEI}$ peak are still missing. A similar procedure for the $O_{carb/hydr}$ peak gives about 8%, which is comparable to the "missing" Li considering typical 1-2 % error in XPS quantification, especially for these rather convoluted spectra.

**Supplementary Note 2: XPS estimation of the CEI thickness after formation cycles**

The thickness of the CEI layer was calculated using the method proposed by Malmgrem et al, based on the decrease of the bulk signal after covering with this over-layer upon cycling [2]. In particular, the thickness $t$ is related to the area ratio of the $O_{latt}$ peak before and after cycling by the following equation:

$$t = \langle\lambda\rangle \sin\theta \ln\frac{A_0}{A}$$

Where $\langle\lambda\rangle$ is the IMFP of the CEI, that we took as 3.9 nm using the TPP2M formula for LiF and kinetic energy of 1116 eV (530 eV measured with the Al X-ray source), $\theta$ the take-off angle of 45°, and A0 and A area the peak areas of the pristine and cycled electrodes, respectively.

We used unscratched electrodes for this calculation due to intensity changes given by the otherwise necessary scratching operation. Note indeed that without this operation, the Ni 2p spectra is hidden by the F KLL Auger lines of PVDF (**Supplementary Figure 6a**). A binding energy correction to the $CF_2$ peak at 290.2 eV was also performed due to the lower conductivity given by PVDF. As shown in **Supplementary Figure 6b**, the $O_{latt}$ peak at ~529 eV was still visible. For these samples, the O 1s peak model discussed above and in the main text was modified to include the P-O bonds at ~534 eV given by phosphate species; the $O_{surf}$ peak was omitted because of the low intensity of $O_{latt}$. The calculated thicknesses was 1.6 for the samples retrieved at 3.2 V. On average, we concluded that the thickness was of order of 1-2 nm.

Below the aforementioned 1-2 nm thick surface layer, O 1s and Ni $2p_{3/2}$ XPS and HAXPES analysis indicate the presence of a surface layer with a reduced state of nickel ions. Visual inspection of the Ni $2p_{3/2}$ spectra (**Figure 4b-e**), combined with estimated sampling depth for each acquisition mode (**Figure 1**) readily suggest that the thickness of such layer would be confined within the first 10 nm (i.e. sampling depth for the measurement at 2.3 keV). However, the spectral shape and fits for the surface-sensitive Al Kα XPS (sampling depth of ~3 nm) measurements already indicate a minority contribution from oxidized $Ni^{III/IV}$. Either a non-uniform or a non-homogeneous surface layer, or a combination of both might explain this effect.

As first approximation, the surface layer thickness can be estimated by assuming a simple homogeneous and uniform model system, which lead to the following formula:

$$t = \lambda \sin\theta \ln(A_{bulk})$$

where $\langle\lambda\rangle$ is the inelastic mean free path ($LiNiO_2$ was assumed), $\theta$ the take off angle, and $A_{bulk}$ the area percentage of the Ni $2p_{3/2}$ peak related to bulk material (i.e. $Ni^{III/IV}$), as estimated by reference-based peak fitting.

However, as shown in **Supplementary Table 5**, this result in a thickness that is photon-energy dependent, which highlights the influence of inhomogeneity. Therefore, XPS thickens estimation is non-trivial, as it requires to consider at the same time the exponential depth distribution of the photoelectron intensity and the layer inhomogeneity. In this framework, the value of ~10 nm given by the most-bulk sensitive HAXPES measurement represent the best estimate as it allows at least to probe deeper within the inhomogeneous surface layer.

To corroborate this conclusion, STEM-EELS investigation was carried out for the most de-intercalated sample ($Li_{0.01}NiO_2$), for which the contrast between surface and bulk is enhanced. As reported in the literature, the oxidation state of Ni can be qualitatively determined by the position of maximum intensity for the Ni $L_{2,3}$-edge EELS spectrum, which shifts to higher energy going from $Ni^{II/III}$ to $Ni^{IV}$ [3,4]. A similar behavior is typically observed in soft XAS spectra, too [5].

Accordingly, a gradual shift is indeed observed for the maximum position of the Ni $L_3$ EELS scans shown in **Supplementary Figure 7a** moving inward the particle with a step of 1 nm. The peak maximum position plotted in **Supplementary Figure 7c** indicate that a plateau is reached after about 10 nm, in fair agreement with HAXPES conclusions.

**Supplementary Note 3: Quantification of O/Ni ratio by Ni 2p and O 1s peak fitting**

The O/Ni at% ratio was quantified by lab-based XPS and HAXPES analysis of the O 1s and Ni 2p core peaks. The surface and bulk contribution were discriminated by O 1s and Ni $2p_{3/2}$ peak fitting. To benchmark the accuracy of the method, commercial-grade NiO and $Ni(OH)_2$ (Sigma-Aldrich) were also analyzed with the same method. To account for the low-binding energy tailing of the main peak, a LF(1,1,70,350) lineshape was found to be more appropriate; standard GL(30) lineshapes were adopted for the other peaks. **Supplementary Figure 8** show the O 1s and Ni 2p spectra highlighting the surface and bulk contributions as light and dark green filled areas. Fit results are shown in **Supplementary Table 7**. Other components in the O 1s spectra related to surface contaminants were not included. For NiO and $Ni(OH)_2$, the whole Ni 2p spectra was considered.

## Supplementary Note 4: Low energy models used for cRPA calculation of the screened Coulomb interaction

To calculate the partial screened Coulomb interaction U(w) for the Ni 3d and O 2p shells, low-energy model Hamiltonians were derived based on ab-initio DFT calculations in the PBE approximation for the end-members $LiNiO_2$ and $NiO_2$. The DFT band structure obtained by non-spin polarized energy and charge self-consistent calculations for NiO, $LiNiO_2$ and $NiO_2$ are shown in **Supplementary Figure 9a,c,e**. In agreement with previous DFT studies, metallic ground states were obtained for NiO and $LiNiO_2$, while $NiO_2$ has a small gap of about 1 eV [6–8]. Experimentally however, none of the three compounds is metallic with either large (NiO, 4-5 eV) or small ($Li_xNiO_2$, <1 eV) band gaps, which highlights the need to treat explicitly electronic correlations within either DFT+U or many-body techniques [9–12].

The Kohn-Sham bands within the energy range from -8 (-10 for NiO) to 4 eV were used to construct maximally localized Wannier functions (MLWFs). The reciprocal overlap integrals define the *pd* low-energy tight binding Hamiltonian whose band structure perfectly matches with the starting DFT representation as shown in **Supplementary Figure** 9**a,c,e** . The density of states projected onto Ni 3d and O 2p MLWFs highlight the strongly hybridized character of the Ni 3d and O 2p bands **Supplementary Figure 9d,f**. This increases from $LiNiO_2$ to $NiO_2$ as observed in the larger O 2p character near the fermi level and the merging of the narrow "$t_{2g}$-like" band with the broad "O 2p-like" band.

The screened Coulomb interaction $U$ is calculated by computing the bare Coulomb interaction $v$ and the partial polarization $P_{partial}$[13]:

$$U^{-1} = v^{-1} - P_{partial}\,, \qquad P_{partial} = P_{total} - P_{corr}$$

where $P_{total}$ and $P_{corr}$ are the total polarization and the contribution from the correlated bands. These are the target states for which the method allows to compute the partial Coulomb interaction. In fact, within the cRPA scheme, $P_{corr}$ is removed by excluding all transitions within the correlated subspace. Therefore, one can see $U$ as the bare Coulomb interaction within the correlated low-energy subspace, which in turn depends on the total polarization due to the whole electronic structure of the material i.e. it is partially screened by channels out of such subspace.

Since the polarization is frequency dependent, U also depends on the frequency $\omega$. The $U(\omega)$ matrix is then evaluated within the set of localized Wannier orbitals $\phi_m$ (m indicating position and angular moment) leading to the $U_{ijkl}(\omega)$ matrix elements [13,14]:

$$U_{ijkl}(\omega) = \langle \phi_i \phi_j | U(\omega) | \phi_k \phi_l \rangle$$

Since both Ni 3d and O 2p bands are included in the definition of Wannier orbitals hence the correlated subspace because of the large pd hybridization in $LiNiO_2$ and $NiO_2$. Therefore, the choice of the target bands to be excluded from the calculation of $P_{partial}$ can include either one of two orbital sets or both of them. Depending on the choice of the model, electronic transitions constituting screening channels to the Coulomb interaction can be switched on or off, as illustrated in **Supplementary Figure 10**. Following nomenclature in the field, we refer to *d-dp* as the model in which only the Ni 3d to Ni 3d transitions are excluded within the *dp* window, while the *dp-dp* model is defined by excluding all reciprocal Ni 3d and O 2p transitions. As a consequence, the main difference between the two models is that the latter does exclude O 2p to Ni 3d transitions.

# Supplementary Figures

## Supplementary Figure 1

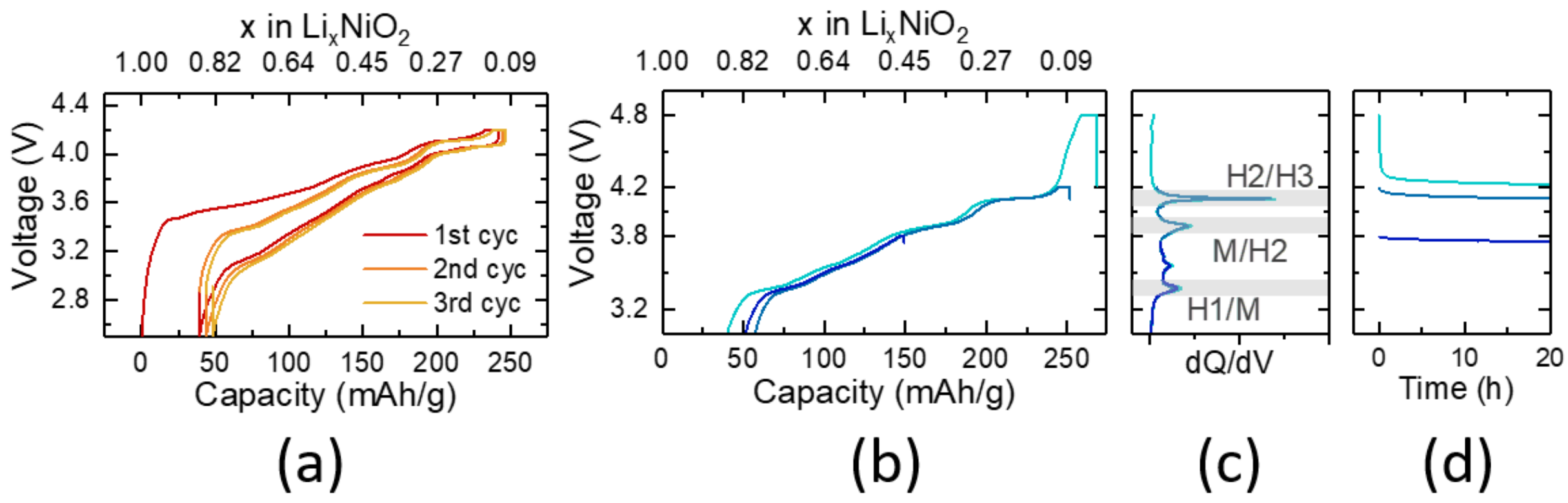

**Supplementary Figure 1. Electrochemical delithiation of $Li_xNiO_2$.** (a) Formation cycles at C/10 for a reference electrode. Apart for a first cycle capacity loss of ~50 mAh/g, no significant changes in the voltage curve were observed. (b) Last charge at C/20 stopped at 3.8, 4.2, and 4.8 V and (c) relative differential capacity curves, highlighting in gray the two-phase transition regions. (d) Voltage decay during the final OCV step, which indicates a faster voltage decay with increasing upper cut-off voltage. This self-discharge process was related to the secondary phases observed by XRD and Raman spectroscopy.

## Supplementary Figure 2

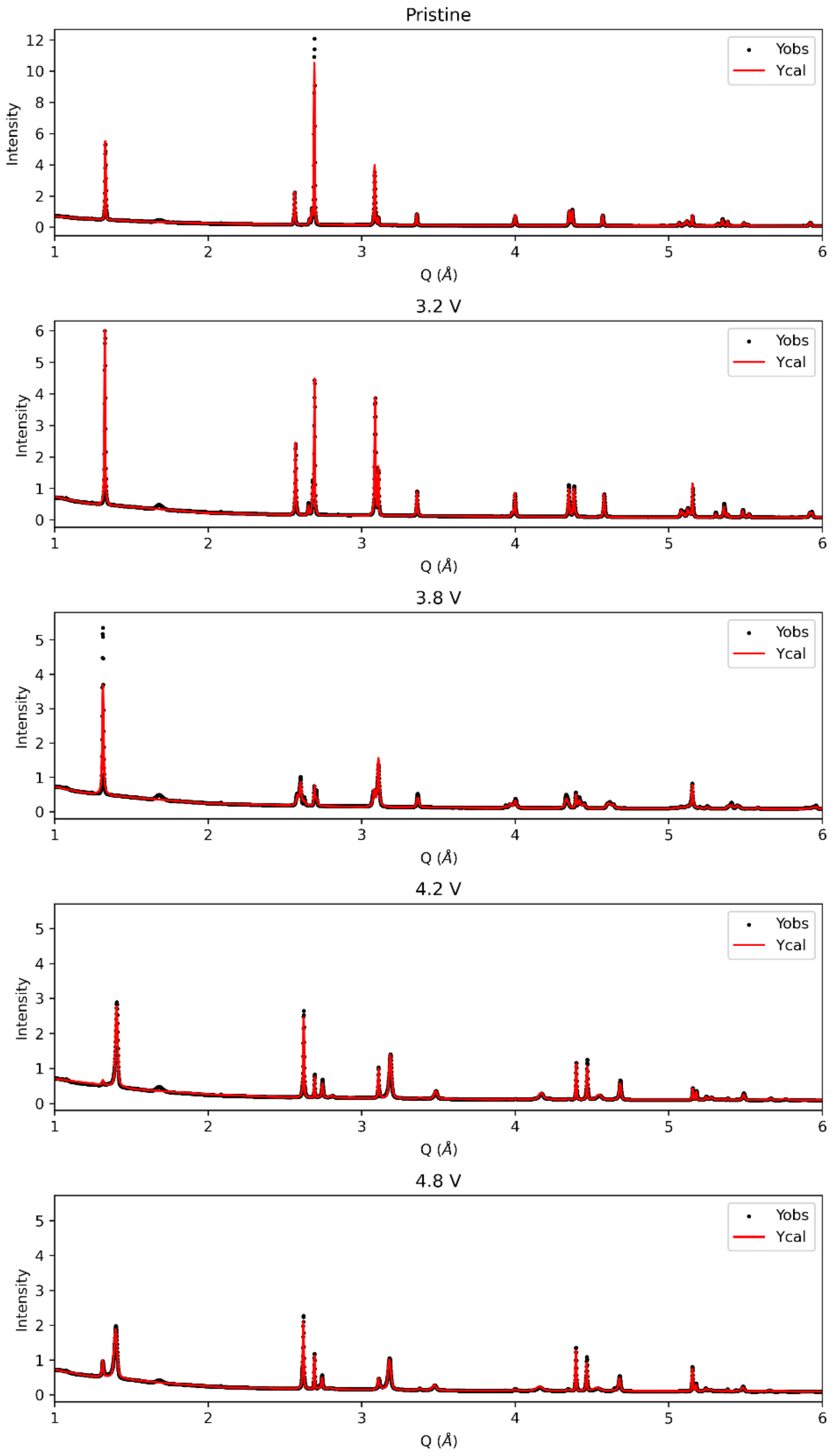

**Supplementary Figure 2.** ***Ex-situ*** **crystal structure characterization of the cycled samples by synchrotron XRD.** Experimental data and refinements are shown as black dots and red lines, respectively. From bottom to top: Pristine, 3.2, 3.8, 4.2 and 4.8 V.

**Supplementary Figure 3**

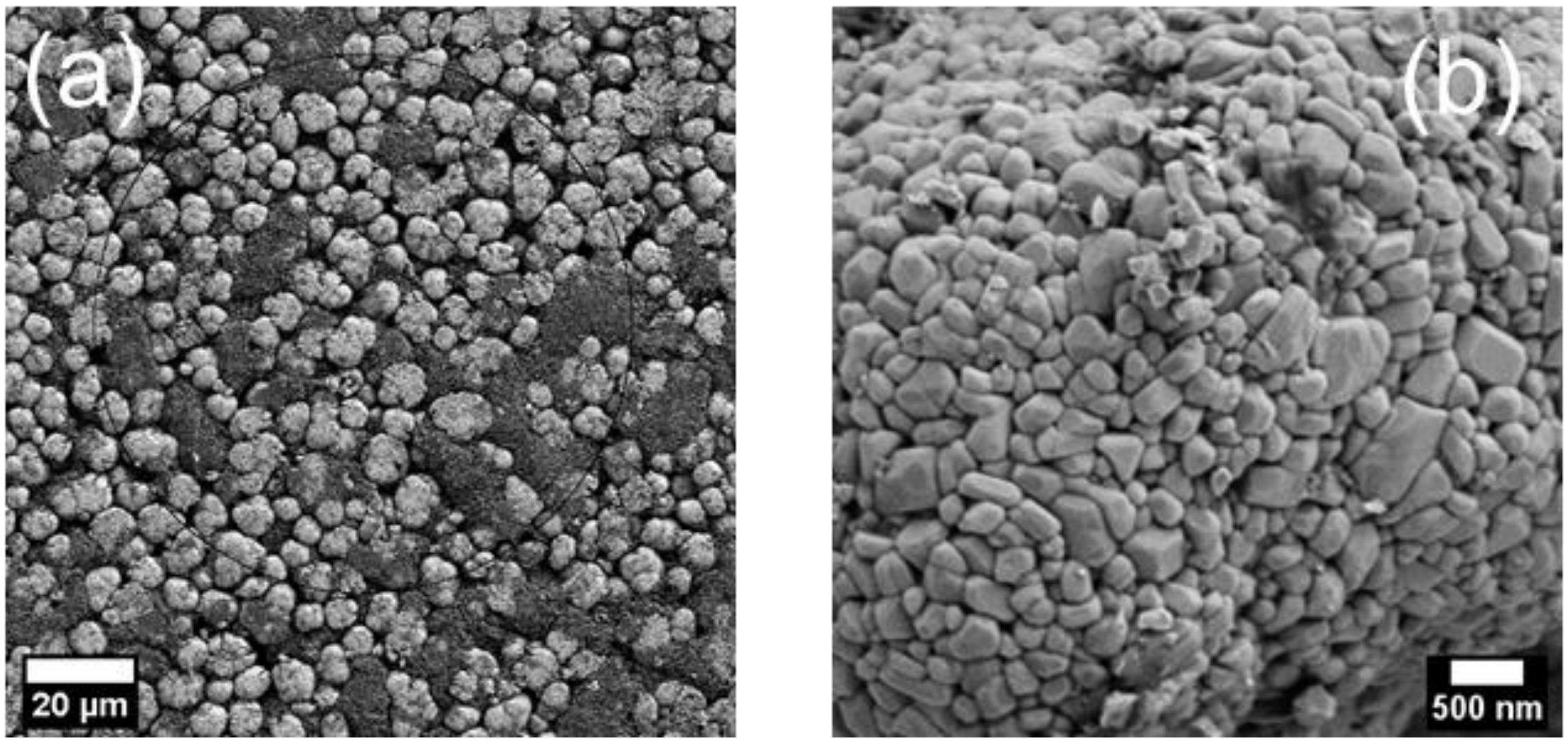

**Supplementary Figure 3. Morphology and particle size analysis of $LiNiO_2$ particles. (a)** Low and (b) high magnification SEM images for the pristine $LiNiO_2$ electrode. As typical commercial-grade Ni-rich electrodes, the morphology of $LiNiO_2$ consists of spherical aggregates with diameter d = 6.9 ± 1.7 µm (average over ~100 particles in **Supplementary Figure 3. Morphology and particle size analysis of LiNiO2 particles. (a)** Low and (b) high magnification SEM images for the pristine LiNiO2 electrode. (b) that are constituted of smaller primary particles with irregular and prismatic-like shapes randomly oriented. The average dimension was estimated as 300 - 400 nm considering ~50 single particles in four different aggregates. Approximating the particles as spheres gives a radius of about 150 - 200 nm, which was compared to XPS and HAXPES depth sensitivities in Figure 1 of the main text.

**Supplementary Figure 4**

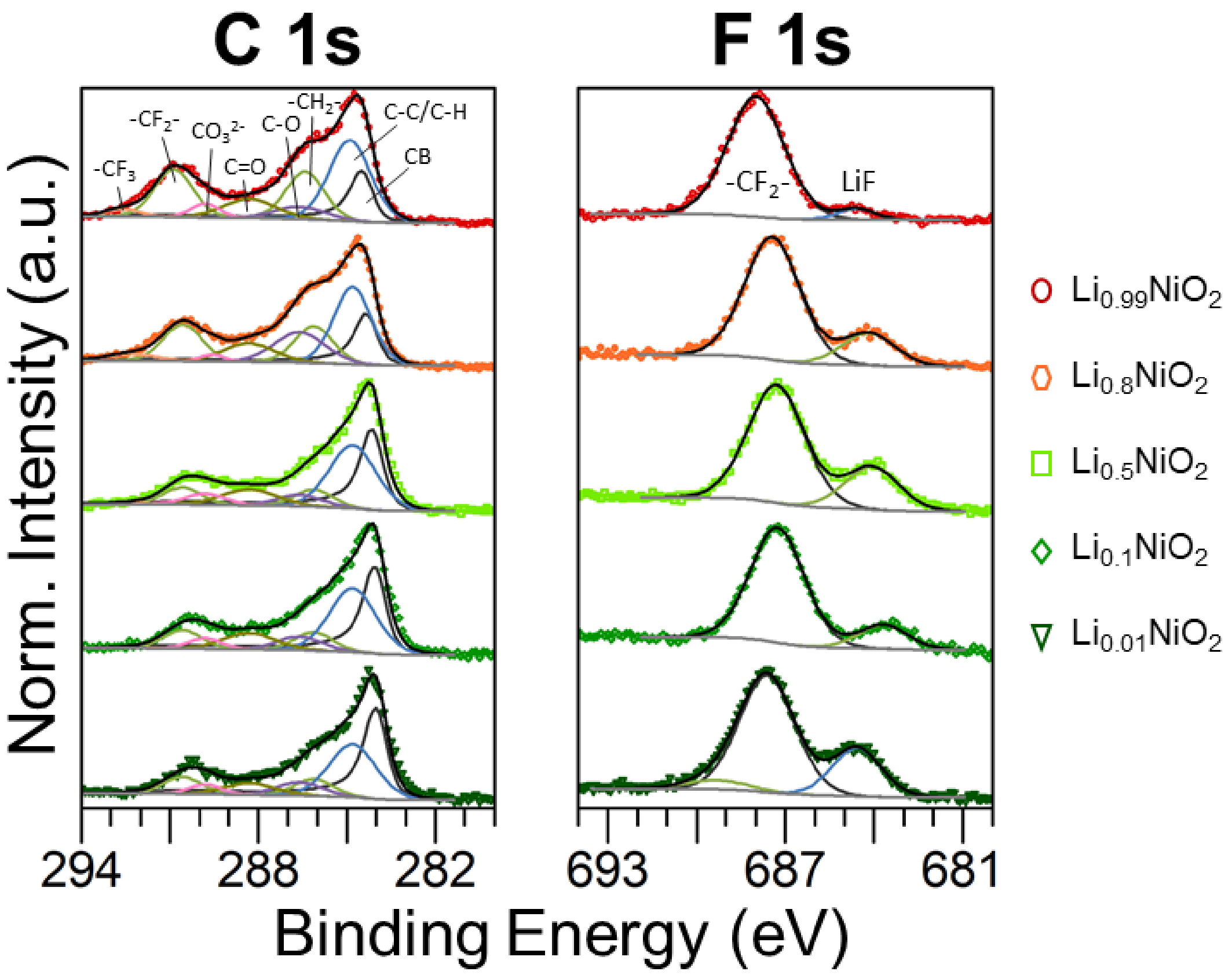

**Supplementary Figure 4. Lab-based XPS C 1s (left) and F 1s (right) core level spectra for the series of $Li_xNiO_2$ electrodes studied in the main text.** All spectra were normalized to its peak maximum. Experimental data is indicated as empty symbols while peak convolution is shown as black line. Peak assignation is indicated in the top spectra.

**Supplementary Figure 5**

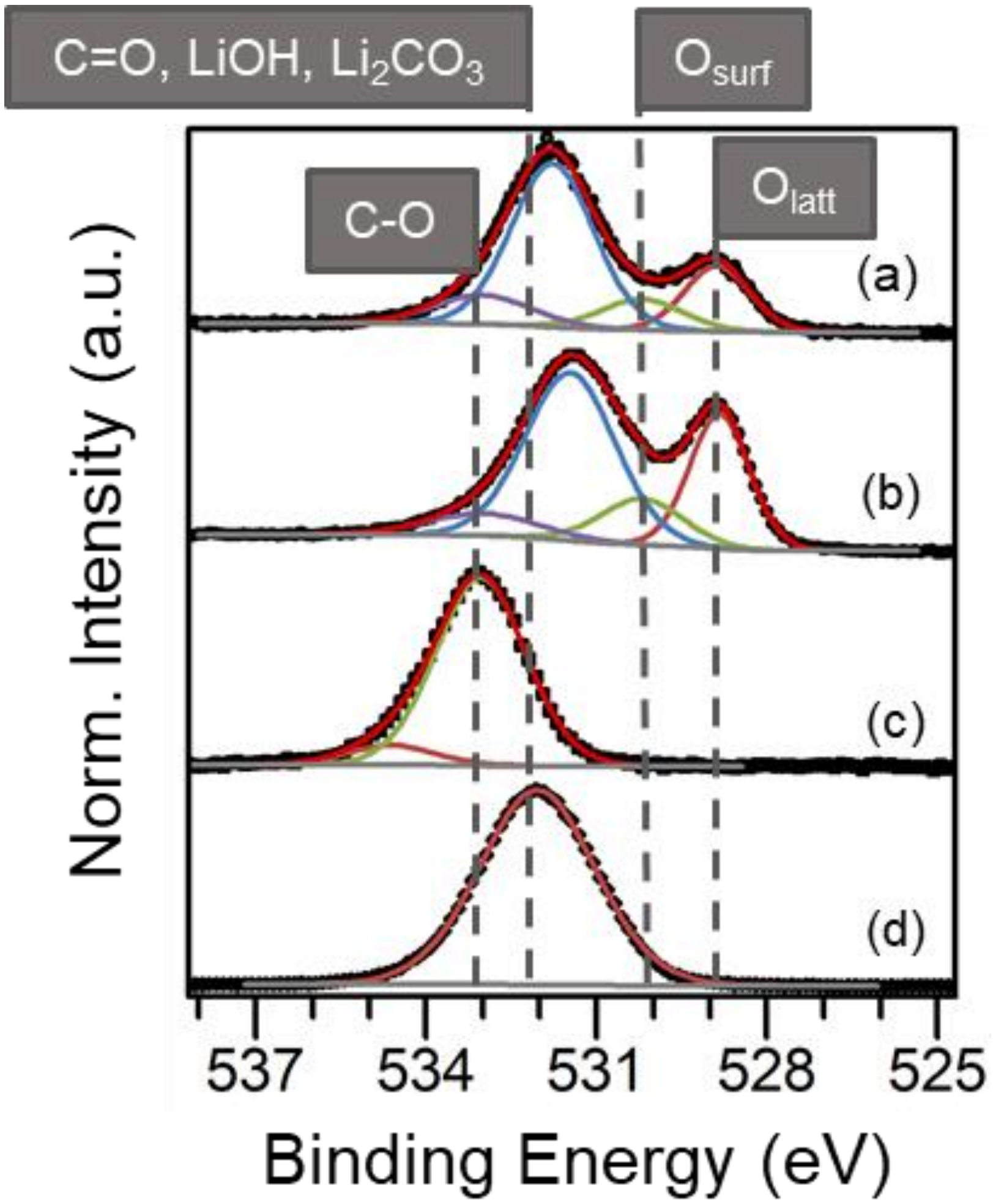

**Supplementary Figure 5. Reference O 1s XPS core level spectra.** Lab-based XPS O 1s core level spectra for (a) pristine electrode, (b) bare LiNiO2 powder, (c) bare PVDF powder, and (c) $Li_2CO_3$ reference sample.

## Supplementary Figure 6

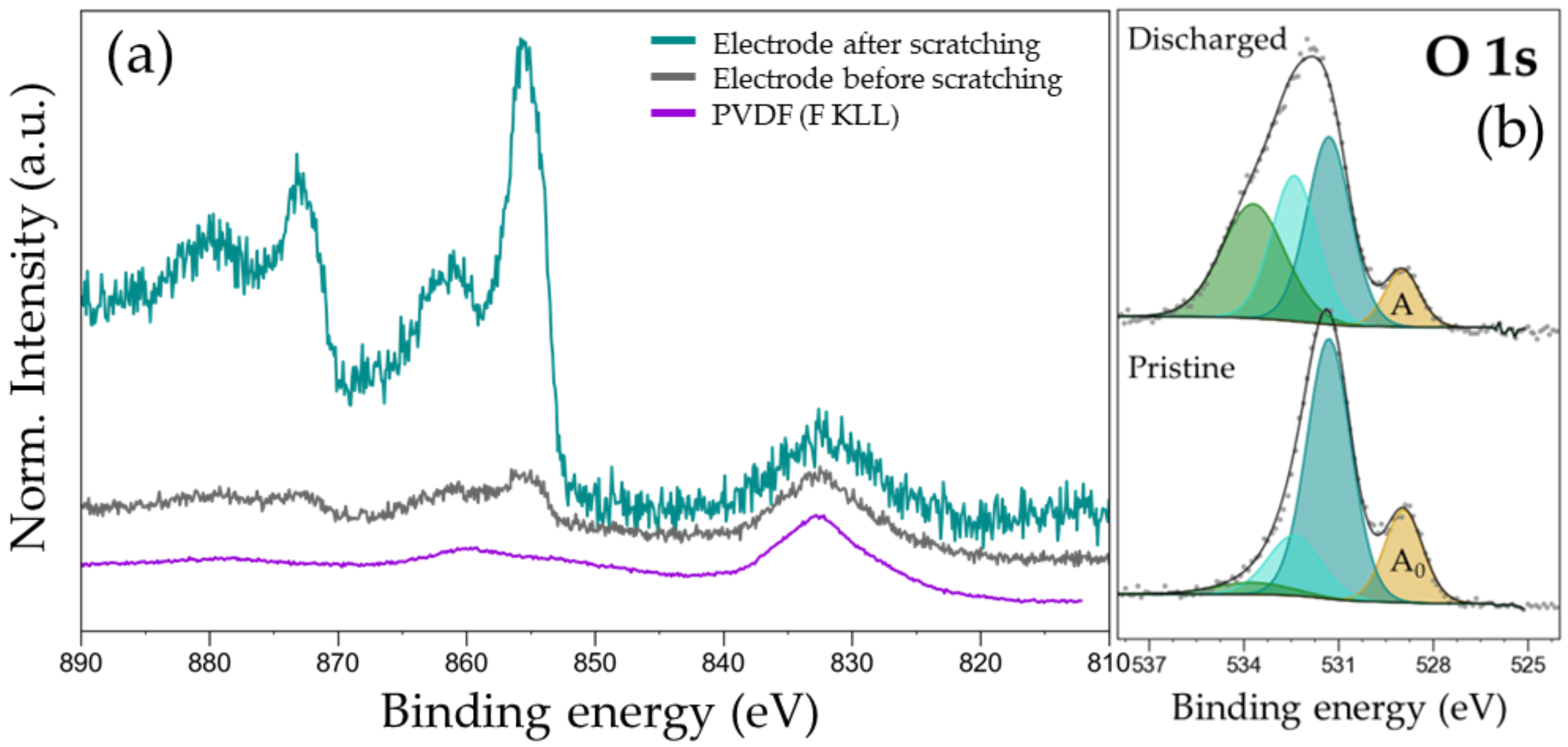

**Supplementary Figure 6. XPS study of unscratched electrodes**. (a) Comparison of Ni 2p spectra before and after scratching, indicating the overlap with F KLL Auger lines that can be overcame by mechanical scratching. The spectra are normalized to the F KLL peak at ~830 eV. (b) O 1s core-level spectra of the unscratched pristine and discharged electrodes.

**Supplementary Figure 7**

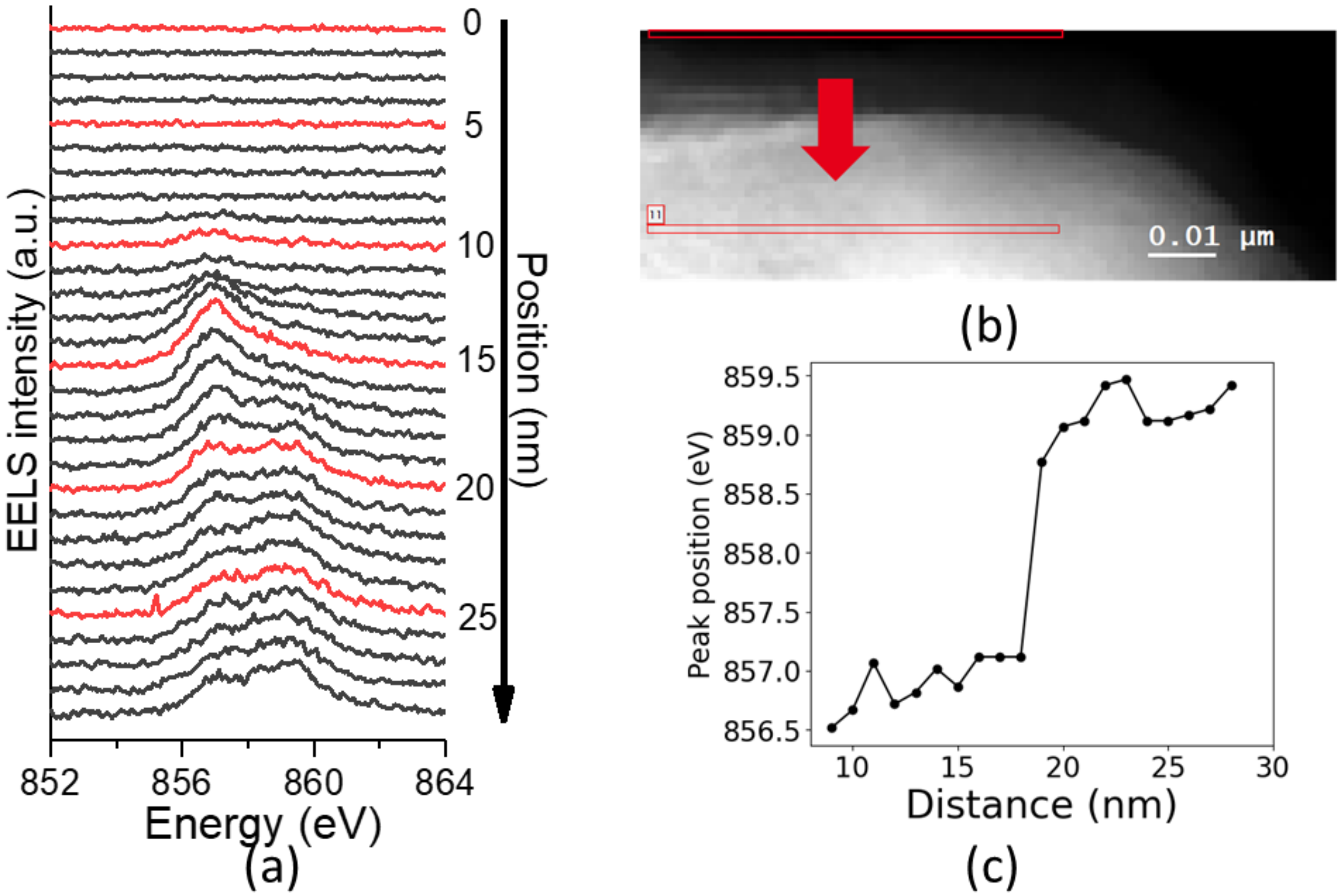

**Supplementary Figure 7. STEM-EELS analysis of the surface reduced layer of the $Li_{0.01}NiO_2$ sample.** (a) High-resolution Ni $L_3$-edge EELS spectra acquired from the extreme surface towards the intern of a particle. (b) Corresponding STEM image showing the first and last position where the EELS spectra were acquired. (c) Energy position of the maxima intensity for each scan; the first measures are not included as no peak was registered.

**Supplementary Figure 8**

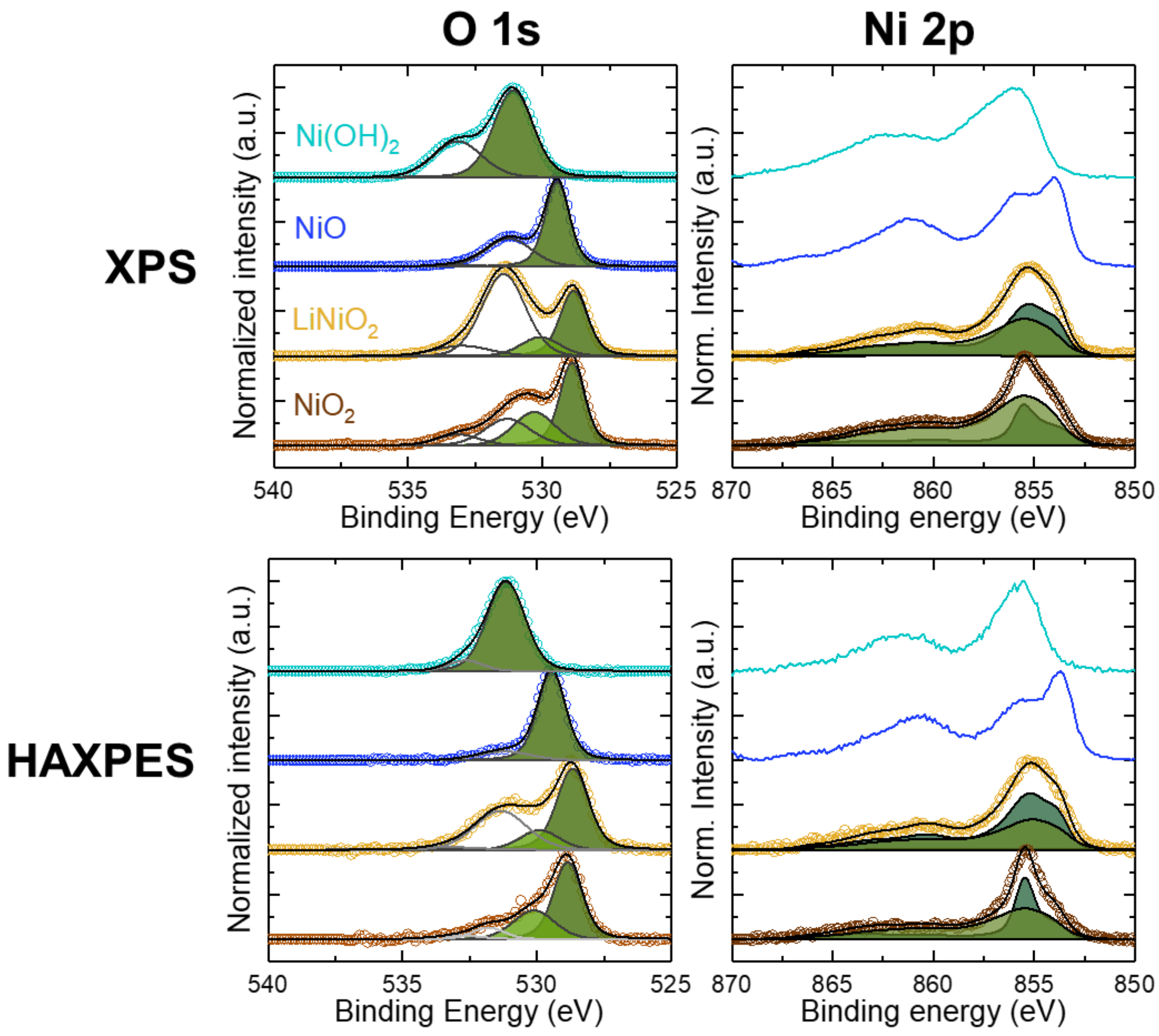

**Supplementary Figure 8 Quantification of the O/Ni atomic ratio by lab-based XPS and HAXPES.** O 1s and Ni $2p_{3/2}$ lab-based XPS and HAXPES spectra for $Li_{0.99}NiO_2$, $Li_{0.01}NiO_2$, $Ni(OH)_2$, and NiO. The highlighted dark and light green peak components were assigned to bulk and surface contribution, respectively.

## Supplementary Figure 9

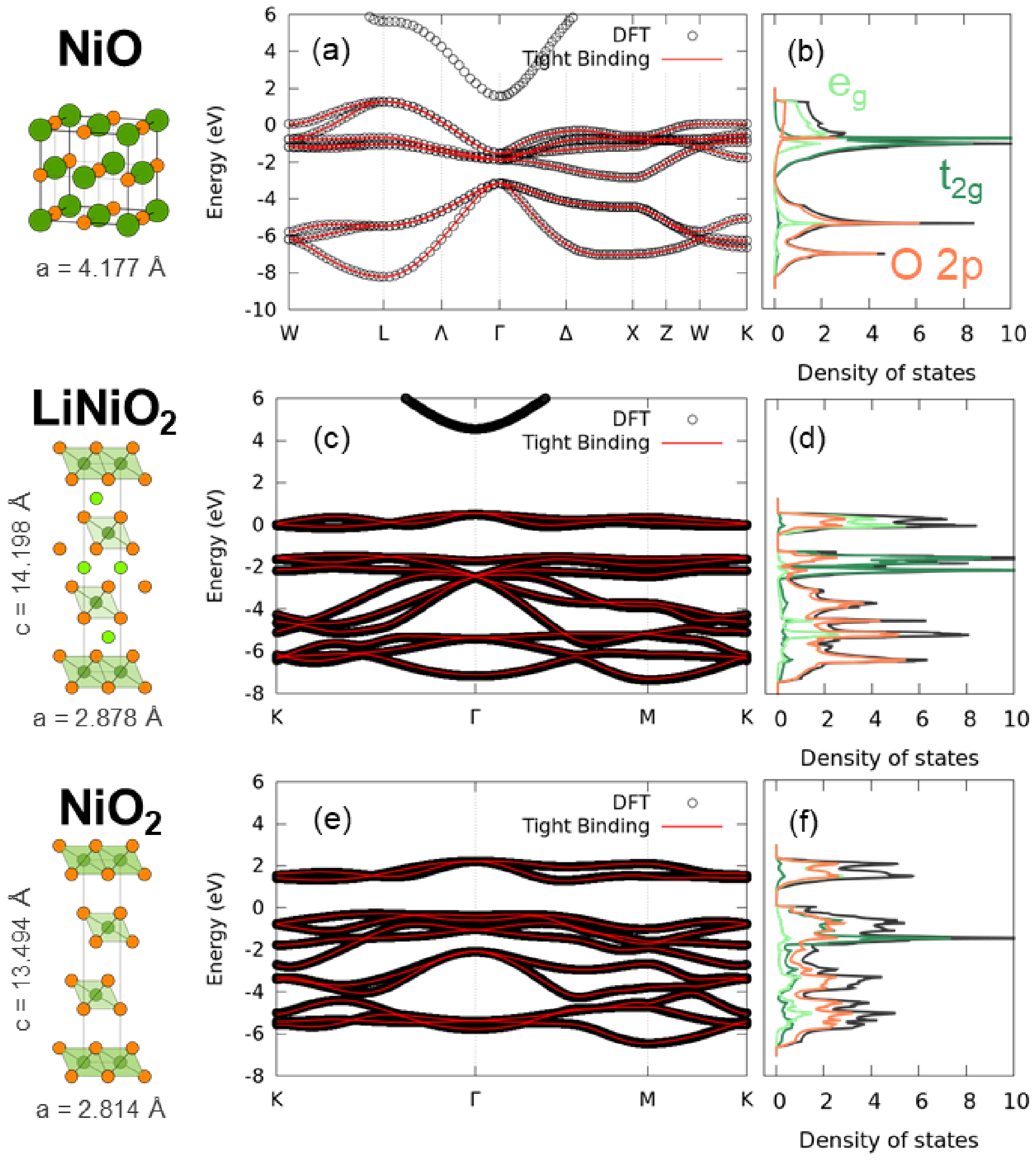

**Supplementary Figure 9. DFT electronic structures used for cRPA and cluster model calculations for NiO, $LiNiO_2$, and $NiO_2$.** The input crystal structure is shown on the left-hand side. Orange, light green and dark green spheres represent oxygen, lithium, and nickel atoms, respectively. (a,c,e) PBE and MLWF tight binding band structures. (b,d,f) total and projected density of states calculated for the MLWFs.

## Supplementary Figure 10

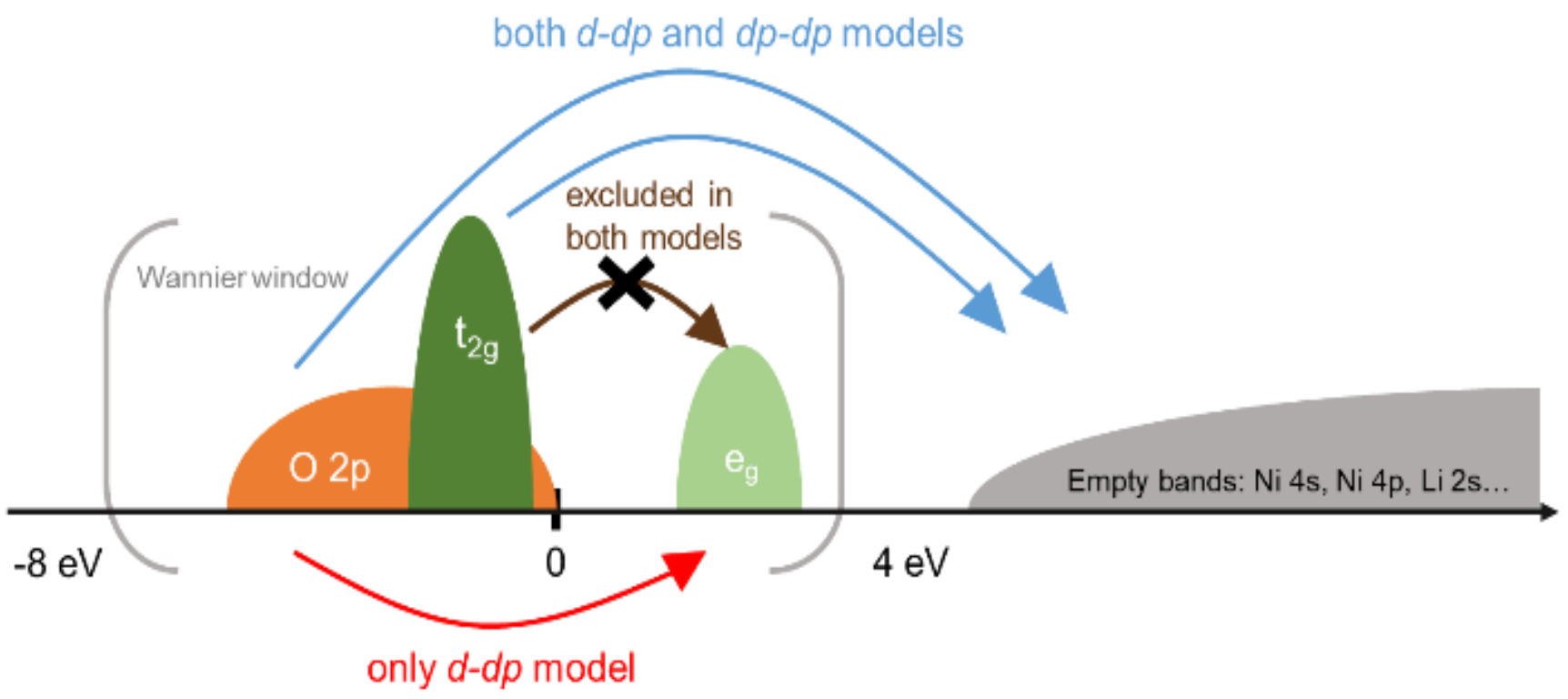

**Supplementary Figure 10. Schematic representation for the screening channels available in the low-energy d-dp and dp-dp models used for the cRPA calculations**. A qualitative density of states for $NiO_2$, approximated from the DFT one, is used as example.

**Supplementary Figure 11**

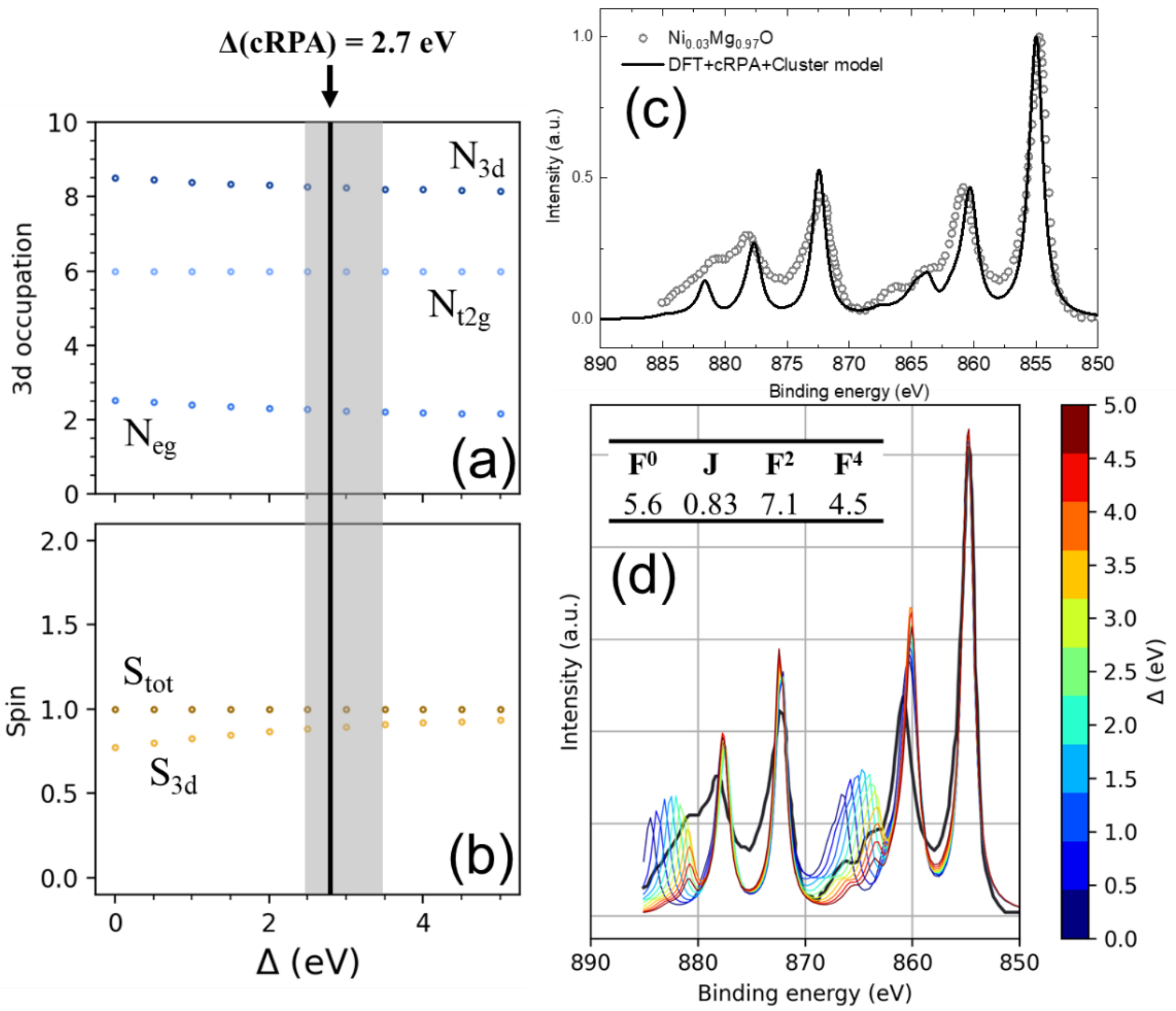

**Supplementary Figure 11. Cluster model analysis of NiO used to test the *ab initio* DFT+cRPA method.** Expectation values for the (a) occupation of the Ni 3d shells and (b) the spin of the total cluster ($S_{tot}$) and of the Ni 3d shell only ($S_{3d}$), obtained by fitting the charge transfer energy (Δ) to the experiment. The grey area indicate the best agreement values and the black line the value of Δ calculated from the cRPA interaction parameters without fitting. Simulated spectra obtained (c) without and (d) with fitting are compared to the experimental Ni 2p spectrum for $Ni_{0.03}Mg_{0.97}O$ taken from Ref. [15].

# Supplementary Tables

## Supplementary Table 1

| Sample | Phase | Space Group | a (Å) | b (Å) | c (Å) | z_O | Phase fraction (%) |
|---|---|---|---|---|---|---|---|
| P | H1 | R -3 m | 2.875 | 2.875 | 14.176 | -0.26 | 100 |
| 3.2 V | H1 | R -3 m | 2.867 | 2.867 | 14.211 | -0.26 | 100 |
| 3.8 V | M | C 2/m | 4.940 | 2.826 | 5.073 | 0.29 | 100 |
| 4.2 V | H3 | R -3 m | 2.813 | 2.813 | 13.439 | -0.26 | 92.9 |
| | H2-like | R -3 m | 2.817 | 2.817 | 14.351 | -0.24 | 7.1 |
| 4.8 V | H3 | R -3 m | 2.814 | 2.814 | 13.494 | -0.26 | 84.5 |
| | H2-like | R -3 m | 2.817 | 2.817 | 14.351 | -0.26 | 15.5 |

**Supplementary Table 2. Parameters obtained by Rietveld refinement of the XRD patterns.** An additional phase for Aluminum (Fm-3m, a = 4.04 Å) was also included in the fits. For the monoclinic phase of sample 3.8 V, β = 109.35 °. For the samples charged up to 4.2 V and 4.8 V, evidence of a secondary "H2-like" phase from the low angle peak was related to the self-discharge occurring upon holding the cut-off voltage and relaxation [16–18]. This process is to be distinguished to the surface degradation analyzed by XPS/HAXPES and observed in all cycled samples as discussed in the main text.

## Supplementary Table 2

| Core level | BE (eV) | λ (nm) | | | |
|---|---|---|---|---|---|
| | | 1.5 keV | 2.3 keV | 5.4 keV | 9.5 keV |
| **Li1s** | 54 | 2.6 | 3.7 | 7.6 | 12.2 |
| **Ni 3p** | 68 | 2.7 | 3.7 | 7.6 | 12.2 |
| **Ni 3s** | 112 | 2.5 | 3.6 | 7.5 | 12.2 |
| **C 1s** | 285 | 2.3 | 3.4 | 7.3 | 12.0 |
| **O 1s** | 529 | 1.9 | 3.1 | 7.0 | 11.7 |
| **Ni 2p** | 855 | 1.4 | 2.6 | 6.6 | 11.3 |
| **Ni 2s** | 1010 | 1.1 | 2.4 | 6.4 | 11.2 |
| **Ni1s** | 8332 | | | | 2.2 |

**Supplementary Table 3. Estimated inelastic mean free path for all core levels accessible for $Li_xNiO_2$ and excited at 1.5, 2.3, 5.4, and 9.5 keV.** The following parameters for $LiNiO_2$ were used: density 4.8 g/cm$^3$, 13 valence electrons, molecular weight 97.63 g/mol, band gap 0.4 eV.

**Supplementary Table 3**

| | Assignation | BE (eV) | % At |
|---|---|---|---|
| **Li 1s** | $LiNiO_2$ | 53.8 | 3.7 |
| | LiF, $Li_2CO_3$, LiOH | 55.0 | 9.5 |
| **Ni 3p** | $LiNiO_2$ | 67.5 | 8.4 |
| **C 1s** | Carbon black (CB) | 284.5 | 8.6 |
| | $CF_2$, $CH_2$ | 286.3 | 9.4 |
| | C-C | 284.9 | 11.1 |
| | C-O | 286.6 | 0.5 |
| | C=O | 288.4 | 3.6 |
| | $Li_2CO_3$ | 289.6 | 1.4 |
| **O 1s** | $LiNiO_2$ bulk lattice ($O_{latt}$) | 528.8 | 8.5 |
| | $LiNiO_2$ surface/defects ($O_{surf}$) | 530.2 | 4.0 |
| | C=O, $Li_2CO_3$, LiOH | 531.5 | 15.9 |
| | C-O | 533.0 | 2.7 |
| **F 1s** | $CF_2$ | 687.9 | 11.0 |
| | LiF | 684.4 | 1.7 |

**Supplementary Table 4. Lab-based XPS quantification for the pristine electrode.**

**Supplementary Table 4**

| | Assignation | BE (eV) | %At |
|---|---|---|---|
| **Li 1s** | $LiNiO_2$ | 53.9 | 5.6 |
| | $Li_2CO_3$,LiOH | 55.3 | 15.0 |
| **Ni 3p** | $LiNiO_2$ | 68.2 | 7.6 |
| **C 1s** | C-C | 284.9 | 13.3 |
| | C-O | 286.3 | 3.3 |
| | C=O | 288.1 | 0.4 |
| | $Li_2CO_3$ | 289.9 | 8.2 |
| **O 1s** | $LiNiO_2$ | 528.8 | 8.7 |
| | $LiNiO_2$ (surf) | 530.0 | 5.1 |
| | C=O,$Li_2CO_3$,LiOH | 531.7 | 27.1 |
| | C-O | 532.6 | 5.8 |

**Supplementary Table 5. Lab-based XPS quantification for the pristine bare powder.**

## Supplementary Table 5

| | $Li_{0.8}NiO_2$ | $Li_{0.5}NiO_2$ | $Li_{0.1}NiO_2$ | $Li_{0.01}NiO_2$ |
|---|---|---|---|---|
| **Al Kα** | -- | 1.1 | 1.5 | 1.2 |
| **2.3 keV** | -- | 3.7 | 5.9 | 4.7 |
| **Cr Kα** | 4.0 | 3.9 | 4.8 | 3.9 |
| **5.4 keV** | 10.9 | 1.7 | 3.9 | 3.9 |
| **9.5 keV** | 13.5 | 3.3 | 10.6 | 10.0 |

**Supplementary Table 6. Surface layer thickness calculated by XPS and HAXPES using a bilayer homogeneous model. All values are in nm.**

**Supplementary Table 6**

| | XPS | | | HAXPES | | |
|---|---|---|---|---|---|---|
| | **Component** | **BE (eV)** | **%at** | **Component** | **BE (eV)** | **%at** |
| **NiO** | **O 1s** | | **61** | **O 1s** | | **52** |
| | $O_{latt}$ | 529.5 | 38 | $O_{latt}$ | 529.5 | 45 |
| | $O_{carb/hydr}$ | 531.3 | 22 | $O_{carb/hydr}$ | 531.2 | 8 |
| | $O_{org}$ | 533.3 | 1 | | | |
| | **Ni 2p** | **854.1** | **39** | **Ni 2p** | **853.7** | **48** |
| **$Ni(OH)_2$** | **O 1s** | | **73** | **O 1s** | | **68** |
| | $O_{latt}$ | 531.1 | 49 | $O_{latt}$ | 531.1 | 62 |
| | $O_{org}$ | 533.2 | 24 | $O_{org}$ | 532.6 | 6 |
| | **Ni 2p** | **856.1** | **27** | **Ni 2p** | **855.5** | **32** |
| **Pristine $LiNiO_2$** | **O 1s** | | **79** | **O 1s** | | **73** |
| | $O_{latt}$ | 528.8 | 23 | $O_{latt}$ | 528.7 | 36 |
| | $O_{surf}$ | 530.1 | 8 | $O_{surf}$ | 529.9 | 10 |
| | $O_{carb/hydr}$ | 531.4 | 41 | $O_{carb/hydr}$ | 531.4 | 26 |
| | $O_{org}$ | 533.0 | 6 | $O_{org}$ | 533.4 | 2 |
| | **Ni 2p** | | **21** | **Ni 2p** | | **27** |
| | $Ni^{II}$ | 855.5 | 10 | $Ni^{II}$ | 855.5 | 11 |
| | $Ni^{III}$ | 855.2 | 11 | $Ni^{III}$ | 855.2 | 16 |
| | $Ni^{IV}$ | 855.6 | 0 | $Ni^{IV}$ | 855.5 | 0 |
| **Charged $LiNiO_2$** | **O 1s** | | **69** | **O 1s** | | **60** |
| | $O_{latt}$ | 528.8 | 25 | $O_{latt}$ | 528.7 | 33 |
| | $O_{surf}$ | 530.2 | 17 | $O_{surf}$ | 530.1 | 17 |
| | $O_{carb/hydr}$ | 531.3 | 21 | $O_{carb/hydr}$ | 531.6 | 6 |
| | $O_{org}$ | 533.3 | 6 | $O_{org}$ | 533.1 | 5 |
| | **Ni 2p** | | **31** | **Ni 2p** | | **40** |
| | $Ni^{II}$ | 855.5 | 22 | $Ni^{II}$ | 855.1 | 22 |
| | $Ni^{III}$ | 854.9 | 7 | $Ni^{III}$ | 855.2 | 9 |
| | $Ni^{IV}$ | 855.6 | 3 | $Ni^{IV}$ | 855.5 | 9 |

**Supplementary Table 7. Lab-based XPS and HAXPES quantification results of Ni 2p and O 1s for reference samples and electrodes.**

# Supplementary References

(1) Fantin, R.; Van Roekeghem, A.; Rueff, J.-P.; Benayad, A. Surface Analysis Insight Note: Accounting for X-Ray Beam Damage Effects in Positive Electrode-Electrolyte Interphase Investigations (under Submission).

(2) Malmgren, S.; Ciosek, K.; Hahlin, M.; Gustafsson, T.; Gorgoi, M.; Rensmo, H.; Edström, K. Comparing Anode and Cathode Electrode/Electrolyte Interface Composition and Morphology Using Soft and Hard X-Ray Photoelectron Spectroscopy. *Electrochimica Acta* **2013**, *97*, 23–32. https://doi.org/10.1016/j.electacta.2013.03.010.

(3) Koyama, Y.; Mizoguchi, T.; Ikeno, H.; Tanaka, I. Electronic Structure of Lithium Nickel Oxides by Electron Energy Loss Spectroscopy. *J. Phys. Chem. B* **2005**, *109* (21), 10749–10755. https://doi.org/10.1021/jp050486b.

(4) Lin, F.; Nordlund, D.; Markus, I. M.; Weng, T.-C.; Xin, H. L.; Doeff, M. M. Profiling the Nanoscale Gradient in Stoichiometric Layered Cathode Particles for Lithium-Ion Batteries. *Energy Environ. Sci.* **2014**, *7* (9), 3077. https://doi.org/10.1039/C4EE01400F.

(5) Li, N.; Sallis, S.; Papp, J. K.; Wei, J.; McCloskey, B. D.; Yang, W.; Tong, W. Unraveling the Cationic and Anionic Redox Reactions in a Conventional Layered Oxide Cathode. *ACS Energy Lett.* **2019**, *4* (12), 2836–2842. https://doi.org/10.1021/acsenergylett.9b02147.

(6) Kim, B.; Kim, K.; Kim, S. Quantification of Coulomb Interactions in Layered Lithium and Sodium Battery Cathode Materials. *Phys. Rev. Mater.* **2021**, *5* (3), 035404. https://doi.org/10.1103/PhysRevMaterials.5.035404.

(7) Aydinol, M. K.; Kohan, A. F.; Ceder, G.; Cho, K.; Joannopoulos, J. *Ab Initio* Study of Lithium Intercalation in Metal Oxides and Metal Dichalcogenides. *Phys. Rev. B* **1997**, *56* (3), 1354–1365. https://doi.org/10.1103/PhysRevB.56.1354.

(8) Chakraborty, A.; Dixit, M.; Aurbach, D.; Major, D. T. Predicting Accurate Cathode Properties of Layered Oxide Materials Using the SCAN Meta-GGA Density Functional. *Npj Comput. Mater.* **2018**, *4* (1), 60. https://doi.org/10.1038/s41524-018-0117-4.

(9) Haverkort, M. W.; Zwierzycki, M.; Andersen, O. K. Multiplet Ligand-Field Theory Using Wannier Orbitals. *Phys. Rev. B* **2012**, *85* (16), 165113. https://doi.org/10.1103/PhysRevB.85.165113.

(10) Sawatzky, G. A.; Allen, J. W. Magnitude and Origin of the Band Gap in NiO. *Phys. Rev. Lett.* **1984**, *53* (24), 2339–2342. https://doi.org/10.1103/PhysRevLett.53.2339.

(11) Anisimov, V. I.; Zaanen, J.; Andersen, O. K. Band Theory and Mott Insulators: Hubbard *U* Instead of Stoner *I*. *Phys. Rev. B* **1991**, *44* (3), 943–954. https://doi.org/10.1103/PhysRevB.44.943.

(12) Korotin, Dm. M.; Novoselov, D.; Anisimov, V. I. Paraorbital Ground State of the Trivalent Ni Ion in $LiNiO_2$ from DFT+DMFT Calculations. *Phys. Rev. B* **2019**, *99* (4), 045106. https://doi.org/10.1103/PhysRevB.99.045106.

(13) Vaugier, L. Electronic Structure of Correlated Materials From First Principles: Hubbard Interaction and Hund's Exchange, Ecole Polytechnique Paris Tech, 2011.

(14) Jiang, H.; Blaha, P. G W with Linearized Augmented Plane Waves Extended by High-Energy Local Orbitals. *Phys. Rev. B* **2016**, *93* (11), 115203. https://doi.org/10.1103/PhysRevB.93.115203.

(15) Altieri, S.; Tjeng, L. H.; Tanaka, A.; Sawatzky, G. A. Core-Level x-Ray Photoemission on NiO in the Impurity Limit. *Phys. Rev. B* **2000**, *61* (20), 13403–13409. https://doi.org/10.1103/PhysRevB.61.13403.
(16) Bautista Quisbert, E.; Fauth, F.; Abakumov, A. M.; Blangero, M.; Guignard, M.; Delmas, C. Understanding the High Voltage Behavior of $LiNiO_2$ Through the Electrochemical Properties of the Surface Layer. *Small* **2023**, 2300616. https://doi.org/10.1002/smll.202300616.
(17) Croguennec, L.; Pouillerie, C.; Delmas, C. $NiO_2$ Obtained by Electrochemical Lithium Deintercalation from Lithium Nickelate: Structural Modifications. *J. Electrochem. Soc.* **2000**.
(18) Croguennec, L.; Pouillerie, C.; Mansour, A. N.; Delmas, C. Structural Characterisation of the Highly Deintercalated $Li_xNi_{1.02}O_2$ Phases (with $x \leq 0.30$). *J. Mater. Chem.* **2001**, *11* (1), 131–141. https://doi.org/10.1039/b003377o.